\documentstyle[12pt]{article}
\newcommand{\be}{\begin{equation}}
\newcommand{\ee}{\end{equation}} %\indent}
\newcommand{\eei}{\end{equation}\indent\indent}
\newcommand{\bc}{\begin{center}}
\newcommand{\ec}{\end{center}}
\newcommand{\ber}{\begin{eqnarray}}
\newcommand{\ear}{\end{eqnarray}}
\newcommand{\ba}{\begin{array}}
\newcommand{\ea}{\end{array}}

\def\case#1/#2{\textstyle\frac{#1}{#2} }

\newcommand{\ds}{\displaystyle}
\def\quot{\par\noindent\hangindent 20pt}
\def\mincir{\raise -2.truept\hbox{\rlap{\hbox{$\sim$}}\raise5.truept
\hbox{$<$}\ }}
\def\magcir{\raise -2.truept\hbox{\rlap{\hbox{$\sim$}}\raise5.truept
\hbox{$>$}\ }}
\begin{document}

\begin{titlepage}
\addtolength{\textwidth}{0.2\textwidth}
%\addtolength{\textheight}{0.3\textheight}
\hoffset=-47pt\voffset=-80pt
\title{\sc Post--Newtonian Cosmological Dynamics in Lagrangian coordinates} 

\author{ 
{\sc Sabino Matarrese} ~~~~\& ~~~{\sc David Terranova}      
\\%[3truemm]
\normalsize {\em Dipartimento di Fisica ``Galileo Galilei", }\\
\normalsize {\em Universit\`a di Padova, via Marzolo 8, 35131 Padova, Italy }
\\ \\
%\normalsize {\em Dipartimento di Fisica ``Galileo Galilei", }\\
%\normalsize {\em Universit\`a di Padova, via Marzolo 8, 35131 Padova,
%Italy } 
   } 
%\date{%$\mbox{}$ \vspace*{0.3truecm} \\
%Submitted to {\em MNRAS}
%put the date here,before {
%%%%%%%%%%%%%%%%%%%%%%%%%%%%%%%%%%%%%%%%%%%%%%%%%%%%%%%%%%%%%%%%%%%%%%
\maketitle
%\vspace*{1truecm}
\begin{abstract}

We study the non--linear dynamics of self--gravitating irrotational 
dust in a general relativistic framework, using synchronous and
comoving (i.e. {\em Lagrangian}) coordinates. All the equations are 
written in terms of a single tensor variable, the metric tensor 
of the spatial sections orthogonal to the fluid flow. This treatment allows
an unambiguous expansion in inverse (even) powers of the speed of light. 
To lowest order, the Newtonian approximation -- in Lagrangian form -- 
is derived and written in a transparent way; the corresponding Lagrangian 
Newtonian metric is obtained. Post--Newtonian corrections are then derived 
and their physical meaning clarified. A number of results are obtained:
{\em i)} the master equation of Lagrangian Newtonian dynamics, the
Raychaudhuri equation, can be interpreted as an equation for the evolution of 
the Lagrangian--to--Eulerian Jacobian matrix, complemented by the 
irrotationality constraint; {\em ii)} the Lagrangian spatial metric 
reduces, in the Newtonian limit, to that of Euclidean 3--space written in 
time--dependent curvilinear coordinates, with non--vanishing 
Christoffel symbols, but vanishing spatial curvature (a particular example of 
it is given within the Zel'dovich approximation); 
{\em iii)} a Lagrangian version of the Bernoulli
equation for the evolution of the ``velocity potential" is 
obtained. {\em iv)} The Newtonian and post--Newtonian content of the 
electric and magnetic parts of the Weyl tensor is clarified. 
{\em v)} At the Post--Newtonian level, 
an exact and general formula is derived for gravitational--wave 
emission from non--linear cosmological perturbations; {\em vi)} a 
straightforward application to the anisotropic collapse of homogeneous 
ellipsoids shows that the ratio of these post--Newtonian terms to the Newtonian 
ones tends to diverge at least like the mass density. {\em vii)}
It is argued that a stochastic gravitational--wave background 
is produced by non--linear cosmic structures, with present--day closure 
density $\Omega_{gw} \sim 10^{-5}$ -- $10^{-6}$ on $1$ -- $10$ Mpc scales. 


%\vspace*{0.5truecm} 
\smallskip
%\ec           
\end{abstract}
\thispagestyle{empty}
\end{titlepage}
\section{Introduction}

The gravitational instability of collisionless matter in a cosmological 
framework is usually studied within the Newtonian approximation, 
which basically consists in neglecting terms whose order is higher than the 
first in metric perturbations around a matter--dominated 
Friedmann--Robertson--Walker 
(FRW) background, while keeping non--linear density perturbations. 
This approximation is usually thought to produce accurate 
results in a wide spectrum of cosmological scales, namely on 
scales much larger than the Schwarzschild radius of collapsing bodies and 
much smaller than the Hubble horizon scale, where the peculiar gravitational 
potential $\varphi_g$, divided by the square of the speed of light $c^2$ to 
obtain a dimensionless quantity, keeps much less than unity, while the peculiar 
matter flow never becomes relativistic. 
To be more specific, the Newtonian approximation 
consists in perturbing only the time--time component of the FRW 
metric tensor by an amount $2\varphi_g/c^2$, where $\varphi_g$ 
is related to the matter density fluctuation $\delta$ via the 
cosmological Poisson equation, 
\be
\nabla_x^2 \varphi_g ({\bf x},t) = 4 
\pi G a^2(t) \varrho_b(t) \delta({\bf x}, t) \;, 
\ee
where $\varrho_b$ is the background matter density and 
$a(t)$ the appropriate FRW scale--factor; 
the Laplacian operator $\nabla_x^2$ has been used here with its standard 
meaning of Euclidean space. The fluid dynamics is then 
usually studied in Eulerian coordinates by accounting for mass conservation 
and using the cosmological version of the Euler equation for a 
self--gravitating pressureless fluid -- as long as the 
flow is in the laminar regime -- to close the system. 
To motivate the use of this ``hybrid approximation", which 
deals with perturbations of the matter and the geometry at a different 
perturbative order, one can either formally expand the correct 
equations of General Relativity (GR) in inverse powers of the speed of light
(e.g. Weinberg 1972), or simply notice that the peculiar 
gravitational potential is strongly suppressed with respect to the matter 
perturbation by the square of the ratio of the perturbation scale $\lambda$ to 
the Hubble radius $r_H= c H^{-1}$ ($H$ being the Hubble constant): 
$\varphi_g/c^2 \sim \delta ~(\lambda / r_H)^2$. 

Such a simplified approach, however, already fails in producing an accurate 
description of the trajectories of relativistic 
particles, such as photons. 
Neglecting the relativistic perturbation of the space--space components 
of the metric, which in the so--called longitudinal gauge is just 
$-2\varphi_g/c^2$, would imply a mistake by a factor of two in 
well--known effects such as the Rees--Sciama (1968) and 
gravitational lensing (e.g. Schneider, Ehlers \& Falco 1992), as it would be 
easy to see, by looking at the solution of the eikonal equation. 
In other words, the level of accuracy not only depends on the peculiar 
velocity of the matter producing the spacetime curvature, but also on the 
nature of the particles carrying the signal to the observer. 
Said this way, it may appear that the only relativistic correction required 
to the usual Eulerian Newtonian picture is that of writing the metric tensor 
in the revised, ``weak field", form (e.g. Peebles 1993)
\be 
ds^2 = - \biggl(1 + {2\varphi_g \over c^2} \biggr) ~c^2 dt^2 + a^2(t) 
\biggl(1 - {2\varphi_g \over c^2} \biggr) ~d l^2 \;.
\ee

However, as we are going to show, this is not the whole story. 
It is well--known in fact that 
the gravitational instability of aspherical perturbations (which is the generic 
case) leads to the formation of very anisotropic structures whenever pressure 
gradients can be neglected (Lynden--Bell 1962; Lin, Mestel \& Shu 1965; 
Zel'dovich 1970; Icke 1973; White \& Silk 1979; Shandarin et al. 1995). 
Matter first flows in almost two--dimensional structures called pancakes, 
which then merge and fragment to eventually form one--dimensional filaments 
and point--like clumps. 
During the process of pancake formation the matter density, the shear 
and the tidal field formally become infinite along evanescent 
two--dimensional configurations corresponding to caustics; after this
event a number of highly non--linear phenomena, such as vorticity 
generation by multi--streaming, merging, tidal disruption and 
fragmentation, occur. 
Most of the patology of the caustic formation process, such as the local 
divergence of the density, shear and tide, and the formation of multi--stream 
regions, are just an artifact of extrapolating the pressureless 
fluid approximation beyond 
the point at which pressure gradients and viscosity become important. 
In spite of these limitations, however, it is generally believed that 
the general anisotropy of the collapse configurations, either pancakes or 
filaments, is a generic feature of cosmological structures originated through 
gravitational instability, which would survive even in the presence of a
collisional component. 

This simple observation already shows the inadequacy of the standard Newtonian 
paradigm. According to it, the lowest scale at which the approximation can 
be reasonably applied is set by the amplitude of the gravitational potential 
and is given by the Schwarzschild radius of the collapsing body, which is 
negligibly small for any relevant cosmological mass scale. 
What is completely missing in this criterion is the role of the shear which 
causes the presence of non--scalar contributions to the metric perturbations. 
A non--vanishing shear component is in fact an unavoidable feature of 
realistic cosmological perturbations and affects the dynamics 
in (at least) three ways, all related to non--local effects, i.e. to the 
interaction of a given fluid element with the environment. 

First, at the lowest perturbative order the shear is related to the 
tidal field generated by the surrounding material by a simple proportionality 
law (because of this linear coincidence, in much of the literature 
``shear" and ``tide" are used as synonima). This sort of non--locality, however,
is coded in the initial conditions of each fluid--element through a 
Coulomb--like interaction with arbitrarily distant matter. Because of 
its link with the initial data of each fluid element one can 
consider it as a {\em local} property. The later modification of these 
shear and tidal fields is one of the consequences of the non--linear evolution. 

Second, it is related to a {\em dynamical} tidal induction: the modification 
of the environment forces the fluid element to modify its shape and density. 
In Newtonian gravity, this is an {\em action--at--a--distance} effect, which 
starts to manifest itself in second--order perturbation theory as an 
inverse--Laplacian contribution to the velocity potential (e.g. Catelan et al. 
1995, and references therein). 

Third, and most important here, a non--vanishing shear field leads to the 
generation of a traceless and divergenceless metric perturbation which can be 
understood as gravitational radiation emitted by non--linear perturbations. 
This contribution to the metric perturbations is statistically 
small on cosmologically interesting scales, but it becomes relevant whenever 
anisotropic (with the only exception of exactly one--dimensional) collapse 
takes place. In the Lagrangian picture considered here, such an effect
already arises at the post--Newtonian (PN) level. 

Note that the two latter effects are only detected if one 
allows for non--scalar perturbations in physical quantities. Contrary to a 
widespread belief, in fact, the choice of scalar perturbations in the initial 
conditions is not enough to prevent tensor modes to arise beyond the linear 
regime in a GR treatment. Truly tensor perturbations are dynamically generated 
by the gravitational instability of initially scalar perturbations, 
independently of the initial presence of gravitational waves. 

This point is very clearly displayed in the GR Lagrangian second--order 
perturbative approach. The pioneering work in this field is by 
Tomita, who, back in 1967, calculated the gravitational waves emitted by 
non--linearly evolving scalar perturbations in an Einstein--de Sitter 
background, in the synchronous gauge (Tomita 1967). Matarrese, Pantano \& Saez 
(1994a,b) obtained an equivalent result but with a different formalism
in comoving and synchronous coordinates. 
According to these calculations, a traceless and divergenceless 
contribution to the spatial metric in the synchronous gauge, 
$\pi^\alpha_{~\beta}$ [greek indices label Lagrangian spatial coordinates, 
while capital latin letters will label Eulerian space; 
lower--case latin indices will be used for spacetime coordinates], 
is produced, which, with growing mode initial conditions, obeys the 
inhomogeneous wave--equation 
\be 
\biggl({1 \over c^2} {\partial^2 \over \partial \tau^2} + {4 \over c^2 \tau } 
{\partial \over \partial \tau } - \nabla_q^2 \biggr)  \pi^\alpha_{~\beta} =
- {\tau^4 \over 42} \nabla_q^2 {\cal S}^\alpha_{~\beta} \;, 
\ee 
where $\tau \propto t^{1/3}$ is the conformal time. The non--linear 
source tensor ${\cal S}^\alpha_{~\beta}$ is given in terms of the initial 
peculiar gravitational potential, $\varphi_0 \equiv \varphi_g(t_0)$, by 
\be 
{\cal S}^\alpha_{~\beta} = 
\delta^\alpha_{~\beta} \nabla_q^2 \psi_0 + {\psi_{0,}}^\alpha_{~\beta}
+ 2 \biggl({\varphi_{0,}}^\alpha{_\beta} \nabla_q^2 
\varphi_0 
- {\varphi_{0,}}^\alpha_{~\mu} {\varphi_{0,}}^\mu_{~\beta} \biggr) \;, 
\ee
with 
\be
\nabla_q^2 \psi_0 = - {1 \over 2} 
\biggl( \bigl(\nabla_q^2 \varphi_0 \bigr)^2 - {\varphi_{0,}}^\mu_{~\nu}
{\varphi_{0,}}^\nu_{~\mu} \biggr) \;, 
\ee
where spatial gradients, indicated by greek indices after a comma, are
with respect to the Lagrangian coordinates $q^\alpha$ and indices are 
raised by the Kronecker symbol; finally $\nabla_q^2$ is just the standard 
(i.e. Euclidean) Laplacian in Lagrangian coordinates. 
To get from the above formula a form 
which can be compared with the standard Newtonian 
interpretation, we can expand $\pi^\alpha_{~\beta}$ in powers of $1/c^2$
(as we will see below, the absence of odd powers of the speed of light is a 
characteristic feature of the Lagrangian approach),
$\pi^\alpha_{~\beta} = {\pi^{(N)}}^\alpha_{~\beta} + \frac{1}{c^2}
{\pi^{(PN)}}^\alpha_{~\beta} + {\cal O} \bigl( \frac{1}{c^4} \bigr)$. 
To zeroth order one obtains the Newtonian term 
${\pi^{(N)}}^\alpha_{~\beta} = \frac{\tau^4}{42} {\cal S}^\alpha_{~\beta}$ 
\footnote{Actually, this expression is determined up to a harmonic 
divergenceless and traceless tensor, which can be set to zero if we 
require consistency with the standard Newtonian second--order results.}, 
which includes a non--local and non--causal contribution to the shear 
tensor through derivatives of the potential $\psi_0$ defined above. 
The meaning of this contribution has been discussed by 
Matarrese, Pantano \& Saez (1994a,b), who obtained it by looking at 
perturbation scales much smaller than the Hubble radius; it 
represents the ``relic" of a causal signal which, on 
sub--horizon scales, appears as an instantaneous Newtonian feature. 
To first order in $1/c^2$ one then gets a PN contribution, 
$\nabla_q^2 {\pi^{(PN)}}^\alpha_{~\beta} = \frac{2 \tau^2}{3} 
{\cal S}^\alpha_{~\beta}$; once again the causality of 
this gravitational--wave signal is lost because of the $1/c^2$ expansion. 
In the formalism of (Matarrese, Pantano \& Saez 1994a,b) this PN term 
would be detected as a sub--leading contribution for perturbation scales 
much smaller than the Hubble radius. 
The close relation between the two approximation schemes 
-- inverse powers of the speed of light and powers of the ratio of the 
perturbation to the horizon scale -- also helps in better understanding the 
actual physical meaning of the $1/c^2$ expansion in the Lagrangian picture. 

The latter PN effect will be recovered in Section 4 without any restriction to 
a second--order perturbation treatment. 
A heuristic estimate of the amplitude of this effect in the frame of current
scenarios of cosmological structure formation is reported in Section 5. 
One can also speculate on the possibility to detect the resulting stochastic 
gravitational--wave background, e.g., through the secondary anisotropy it
would induce on the Cosmic Microwave Background (CMB). 

There is an intriguing aspect of the above expression for the tensor modes 
in the PN limit, namely $\nabla_q^2 {\pi^{(PN)}}^\alpha_{~\beta} = 
\frac{2 \tau^2}{3} 
{\cal S}^\alpha_{~\beta}$. Our explicit PN result of Section 4, if further 
approximated to a perturbative second--order expansion around the FRW 
background, gives $\nabla_q^2 {\pi^{(PN)}}^\alpha_{~\beta} = \frac{\tau^2}{9} 
{\cal S}^\alpha_{~\beta}$, i.e. a factor of 6 smaller! Should one
conclude that the $1/c^2$ expansion and that in the amplitude 
of the perturbations around FRW do not commute? The explanation is actually 
much simpler: the splitting of the various perturbation modes into scalars, 
vectors ans tensors is obviously background--dependent; in particular, our PN 
tensor modes are defined with respect to a non--perturbative Newtonian 
background, while the tensor modes obtained in second--order perturbation 
theory are defined with respect to the usual FRW solution. When the PN 
expressions for the various geometric modes are further expanded to second 
order in perturbation theory and then compared to those obtained through a 
second--order followed by a $1/c^2$ expansion the results do not generally 
coincide (because the Newtonian background itself contains perturbations of 
the FRW metric), but their sum, i.e. the overall metric perturbation, does. 
This completely accounts for the different numerical factor. 

Finally, Eq.(3) allows to understand another important point: the complete
insensitivity of the Newtonian approximation to the possible presence of 
free gravitational waves in the initial conditions, 
such as those produced by quantum effects in the 
early universe. These initial tensor modes, 
corresponding to solutions of the homogeneous equation associated to Eq.(3),
would reduce to harmonic, transverse and traceless 
metric perturbations in the Newtonian limit, having no effect on physical 
quantities (they are gauge modes from the point of view of the Newtonian 
equations). 

The reader at this point may be confused by the continuous interchange of 
Newtonian and PN concepts. However, this will appear unavoidable once one 
realizes that, as in any perturbative treatment (the perturbation parameter 
here being formally $1/c^2$), there are equations which mix different 
perturbation orders. So, the PN equations will have Newtonian sources, or read 
the other way around, there are Newtonian effects which are produced by PN 
sources.  
This point has been definitely clarified in a fundamental paper by 
Kofman \& Pogosyan (1995), who showed how the Newtonian ``electric" tidal field 
$E^\alpha_{~\beta}$ evolves in time according to a PN equation, so that the 
circulation of the PN ``magnetic" Weyl tensor $H^\alpha_{~\beta}$, happens to 
be responsible for the Newtonian non--local ``tidal induction". Bertschinger \& 
Hamilton (1994) gave a different interpretation of the same effect. 

Recently a number of different approaches to 
relativistic effects in the non--linear dynamics of cosmological 
perturbations have been proposed. Matarrese, Pantano \& Saez (1993) proposed 
an algorithm based on neglecting the magnetic part of the Weyl tensor 
in the dynamics, obtaining strictly local fluid--flow evolution equations, 
i.e. the so--called ``silent universe". Using this formalism 
Bruni, Matarrese \& Pantano (1995a) studied the asymptotic behaviour of 
the system, both for collapse and expansion, showing, in particular, that this 
kind of local dynamics generically leads to spindle singularities for
collapsing fluid elements, thereby confirming the results of a previous 
analysis by Bertschinger \& Jain (1994). This formalism, however, 
cannot be applied to cosmological structure formation {\em inside} the horizon, 
where the non--local tidal 
induction cannot be neglected, i.e. the magnetic Weyl tensor 
$H^\alpha_{~\beta}$ is non--zero, with the exception of highly specific initial 
configurations (Matarrese, Pantano \& Saez 1994a; Bertschinger \& Jain 1994). 
Rather, it is probably related to the non--linear dynamics of an irrotational 
fluid {\em outside} the (local) horizon (Matarrese, Pantano \& Saez 1994a,b). 
One possible application (Bruni, Matarrese \& Pantano 1995b), is in fact 
connected to the so--called {\em Cosmic No--hair Theorem} 
(e.g. Hawking \& Moss 1982), 
i.e. to the conjecture that expanding patches of an initially inhomogeneous 
and anisotropic universe asymptotically tend to almost FRW solutions, thanks to
the action of a cosmological constant--like term. 
The self--consistency of these ``silent universe" models has been recently 
demonstrated by Lesame, Dunsby \& Ellis (1995), extending an earlier analysis 
by Barnes \& Rowlingson (1989). Lesame, Ellis \& Dunsby (1996) discussed the 
role of the divergence of the magnetic Weyl tensor\footnote{Actually, 
a non--zero ${\rm div} H$ only appears as a third--order effect in the 
amplitude of perturbations around FRW, unless the initial conditions contain a 
mixture of growing and decaying modes.}, whose presence 
reflects the fact that the shear and the electric tide generally have a 
different eigenframe. 
A local--tide approximation for the non--linear evolution of collisionless 
matter, which tries to overcome some limitations of the 
Zel'dovich approximation (Zel'dovich 1970), has been recently proposed by 
Hui \& Bertschinger (1995). 

In this work we will follow the more ``conservative" approach of 
expanding the Einstein and continuity equations in inverse powers of the 
speed of light, which will then define a Newtonian limit and, at the next 
order, post--Newtonian corrections. The newer aspect of our approach 
is the choice of gauge: we use synchronous and comoving coordinates, because
of which our approach can be legitimately called a Lagrangian one. Thanks 
to this choice, the dynamical variables involved are quite different to the 
standard ones; the gravitational potential, for instance, never appears 
explicitly in our expansion.

Various approaches have been proposed in the literature, which are somehow 
related to the present one. A PN approximation has been followed by Futamase 
(1988, 1989, 1991) to describe the dynamics of a clumpy universe;
he however used non--comoving coordinates and focused his analysis 
on applications related to the so--called averaging problem in cosmology 
(e.g. Ellis 1984). Tomita (1988, 1991) also used non--comoving coordinates in
a PN approach to cosmological perturbations. 
Shibata \& Asada (1995) recently developed a PN approach to cosmological 
perturbations, but they also used non--comoving coordinates. 
Kasai (1995) [see also (Kasai 1992, 1993)] analyzed the non--linear 
dynamics of dust in the synchronous and comoving gauge; his approximation 
methods are however largely different. 
Finally, in a series of papers, based on the Hamilton--Jacobi approach
(Croudace et al. 1994; Parry et al. 1994; Salopek, Stewart \& Croudace 1994) 
a new approximation technique has been developed, which relies on an expansion 
in higher and higher gradients of an initial perturbation ``seed". In spite of 
its elegance and 
generality, however, this approximation scheme is {\em by construction} unable 
to reproduce the non--local aspects of the gravitational instability on
sub--horizon scales; more specifically, terms containing the inverse of
the Laplacian operator, which are unavoidable in the Newtonian limit, 
would formally require an infinite series of terms in 
a gradient expansion. 

To help the reader from not being too much confused by the various 
perturbative techniques adopted in this work, we anticipate that our 
calculations contain, in different parts, three different kinds of expansion. 

First, the entire paper is mostly based on an expansion in inverse 
even powers of the speed of light: the lowest order -- or background --
solution, in this case, describes the so--called Newtonian approximation 
in an expanding universe. Although in general we do not know
the explicit form of the Newtonian background, we can safely assume it 
exists and use it to derive the next order terms. The result of 
first--order perturbation theory is then called post--Newtonian (PN); 
the second order would be the post--post--Newtonian 
(PPN) approximation. The range of application of this perturbative method 
has been already discussed above. Going to higher and higher orders would 
generally lead to a more accurate description of the system, account 
for some new relativistic effects, such as the generation of gravitational 
waves, and possibly allow for an extension of the range of scales to which the 
formalism can be applied. 

Second, we will also use the most standard cosmological perturbation theory
(see, e.g., Kodama \& Sasaki 1984, and references therein; for 
a pressureless medium, see also Hwang 1994), which is basically an 
expansion in powers of the amplitude of the perturbations around a 
background, homogeneous and isotropic, FRW solution. The first--order, or 
``linear", terms of the expansion are given in Section 2.4. No second--order 
calculations, in this sense, will be presented here, with the exception of 
Eqs.(3) -- (5) reported above. The works by Tomita (1967), 
Matarrese, Pantano \& Saez (1994a,b) and -- in some respect -- Pyne \& Carroll 
(1995) follow precisely this perturbative 
approach up to second order and within GR. The range of applicability of this 
second perturbation technique is that of small fluctuations around a FRW 
background, but with no extra limitations on scale. Going to higher and higher 
orders here generally helps to follow the gravitational instability process 
on a longer time--scale and to account for new non--linear and non--local 
phenomena. 

Third, there is another meaning of ``perturbation theory" in Lagrangian 
coordinates, which is frequently used in the cosmological literature
(e.g. Buchert 1995, and references therein). 
This refers to an expansion, within Newtonian gravity, in powers of the 
displacement vector from Lagrangian to Eulerian coordinates 
(Buchert 1989; Moutarde et al. 1991; Bouchet et al. 1992; 
Buchert 1992; Catelan 1995), 
the background being once more represented by the FRW models. The linear 
result is the so--called Zel'dovich approximation 
(see Section 3.2 below), while the second order terms are either called 
``second--order Lagrangian" or ``post--Zel'dovich"
(e.g. Munshi, Sahni \& Starobinsky 1994). The peculiarity of this 
treatment, at any order, is that, while the displacement vector is
calculated from the equations at the required perturbative order, all the 
other dynamical variables, such as mass density, shear and 
so on, are calculated {\em exactly} from their non--perturbative definition. 
What comes out is a fully non--linear description of the system, which,
though not being generally correct, ``mimics" the true non--linear behaviour. 
This perturbation treatment basically exploits the advantages of the 
Lagrangian picture, leading, in particular, to a more accurate description of 
high density regions. Its limitations are generally set by the emerging of 
caustic singularities, besides those deriving from the underlying Newtonian 
approximation. A similar, Zel'dovich--like, approach can also be followed 
within GR [some progress in this direction has been recently 
made by Kasai (1995)]; this will be however the subject of a future 
investigation. 

The plan of the paper is as follows. In Section 2 we introduce the GR 
Lagrangian formalism. Although we do not use the whole machinery of the 
ADM approach (Arnowitt, Deser \& Misner 1962), some of the language is the 
same; in particular, the clear distinction between constraint and 
evolution equations plays a key role also in our work. 
Section 3 deals with the Newtonian limit of the GR equations in Lagrangian 
coordinates and gives a number of formal applications of the approach. 
Section 4 is instead devoted to the post--Newtonian limit of the GR equations. 
In particular, we discuss the dynamical role of gravitational waves 
generated by non--linear cosmic structures. It is shown that, during 
the collapse of a non--spherical (and non--planar) perturbation, PN 
tensor modes are produced, so as to give a {\em dominant} contribution near 
the collapse time. Conclusions are drawn in Section 5, which also contains 
a qualitative discussion on the amplitude of the PN gravitational--wave modes,
as well as some speculations on their possible detectability. 

%%%%%%%%%%%%%%%%%%%%%%%%%%%%%%%%%%%%%%%%%%%%%%%%%%%%%%%%%%%%%%%%%%%%%%%
\section{Relativistic dynamics of irrotational dust in the Lagrangian
picture}
%%%%%%%%%%%%%%%%%%%%%%%%%%%%%%%%%%%%%%%%%%%%%%%%%%%%%%%%%%%%%%%%%%%%%%%

In this section we will derive the equations governing the evolution of an
irrotational fluid of dust (i.e. $p=\omega=0$) in a synchronous and comoving 
system of coordinates (actually the possibility of making these two gauge 
choices simultaneously is a peculiarity of irrotational dust, which holds at 
any time,  i.e. also beyond the linear regime). 
The starting point will be the Einstein equations 
$R_{ab} - \frac{1}{2} g_{ab} R = \frac{8\pi G}{c^4} T_{ab}$, with 
$R_{ab}$ the Ricci tensor, and the continuity 
equation $T^{ab}_{~~;a}=0$ 
for the matter stress--energy tensor 
$T^{ab}=\varrho c^2 u^a u^b$, where $\varrho$ is the mass density and
$u^a$ the fluid four--velocity (normalized to $u^a u_a =-1$); 
a semicolon denotes covariant differentiation. 
The line--element reads 
\be
ds^2 = - c^2 dt^2 + h_{\alpha\beta}({\bf q}, t) 
dq^\alpha d q^\beta \;. 
\ee
The fluid four--velocity in comoving coordinates is 
$u^a = (1,0,0,0)$. A fundamental quantity of our analysis will be the 
velocity--gradient tensor, which is purely spatial, 
\be
\Theta^\alpha_{~\beta} \equiv c u^\alpha_{~;\beta} = {1 \over 2} 
h^{\alpha\gamma} \dot h_{\gamma\beta} \;, 
\ee
where a dot denotes partial differentiation with respect to the proper time 
$t$. The tensor $\Theta^\alpha_{~\beta}$ represents the {\em extrinsic 
curvature} of the spatial hypersurfaces orthogonal to $u^a$. 

Thanks to the spacetime splitting obtained in our frame, the 10 Einstein 
equations can be immediately divided into 4 constraints and 6 evolution 
equations. The time--time component of the Einstein equations is the 
so--called {\em energy constraint} of the ADM 
approach, which reads 
\be
\Theta^2 - \Theta^\alpha_{~\beta} \Theta^\beta_{~\alpha} + c^2 ~^{(3)}\!R = 
16 \pi G \varrho \;,
\ee
where the {\em volume--expansion scalar} $\Theta$ is just the trace of the 
velocity--gradient tensor, $^{(3)}\!R$ is the trace of the 
three-dimensional Ricci curvature, $^{(3)}\!R^\alpha_{~\beta}$, of the spatial 
hypersurfaces of constant time. 

The space--time components give the {\em momentum constraint}, 
\be 
\Theta^\alpha_{~\beta;\alpha} = \Theta_{,\beta} \;. 
\ee 

Finally, the space--space components represent the only truly 
{\em evolution equations}, i.e. those which contain second--order time 
derivatives of the metric tensor. They indeed govern the evolution 
of the extrinsic curvature tensor and read 
\be 
\dot \Theta^\alpha_{~\beta} + \Theta \Theta^\alpha_{~\beta} + 
c^2 ~^{(3)}\!R^\alpha_{~\beta} = 4 \pi G \varrho \delta^\alpha_{~\beta} \;. 
\ee
Taking the trace of the last equation and combining it with the 
energy constraint, we obtain the {\em Raychaudhuri equation} 
(Raychaudhuri 1957),
\be
\dot \Theta + \Theta^\alpha_{~\beta} \Theta^\beta_{~\alpha} + 4 \pi G \varrho
= 0 \;. 
\ee
Mass conservation is provided by the equation
\be 
\dot \varrho = - \Theta \varrho \;. 
\ee

Given that $\Theta = {1 \over 2} h^{\alpha\gamma} \dot h_{\gamma\alpha} =
\partial (\ln h^{1/2}) / \partial t$, where $h \equiv {\rm det} 
~h_{\alpha\beta}$, we can write the solution of this equation in the form 
\be
\varrho({\bf q}, t) = \varrho_0({\bf q}) \bigl[h({\bf q}, t)/ 
h_0 ({\bf q}) \bigr]^{-1/2} \;. 
\ee
Here and in what follows quantities with a subscript $0$ are meant to be 
evaluated at some initial time $t_0$. 

Finally, let us introduce the so--called {\em electric} and {\em magnetic}
parts of the Weyl tensor, which are both symmetric, flow--orthogonal and 
traceless. They read, respectively,
\be
E^\alpha_{~\beta} = {1 \over 3} \delta^\alpha_{~\beta} \biggl( 
\Theta^\mu_{~\nu} \Theta^\nu_{~\mu} - \Theta^2 \biggr) + \Theta 
\Theta^\alpha_{~\beta} - \Theta^\alpha_{~\gamma} \Theta^\gamma_{~\beta} 
+ c^2 \biggl( ~^{(3)}\!R^\alpha_{~\beta} - {1 \over 3} \delta^\alpha_{~\beta} 
~^{(3)}\!R \biggr)
\ee
and 
\be 
H^\alpha_{~\beta} = {1 \over 2} h_{\beta\mu} \biggl( \eta^{\mu\gamma\delta}
\Theta^\alpha_{~\gamma;\delta} + \eta^{\alpha\gamma\delta}
\Theta^\mu_{~\gamma;\delta} \biggr) \;,
\ee
where $\eta^{\alpha\beta\gamma} = h^{-1/2} \epsilon^{\alpha\beta\gamma}$ is the 
three--dimensional, completely anti--symmetric Levi--Civita tensor 
relative to the spatial metric $h_{\alpha\beta}$ and 
$\epsilon^{\alpha\beta\gamma}$ is such that $\epsilon^{123}=1$, etc... .

Notice that, while the definition of the electric tide $E^\alpha_{~\beta}$ 
is completely fixed, because of its well--known Newtonian limit, 
the magnetic tensor field has no straightforward Newtonian counterpart, and 
can be therefore defined up to arbitrary powers of the speed of light. 
The definition we are adopting here is the most straightforward one; it 
is such that no explicit powers of $c$ appear 
in Eq.(15), which means that its 
physical dimensions are $1/c$ those of $E^\alpha_{~\beta}$. This choice 
can be motivated in analogy with electrodynamics, where the 
magnetic vector field is also scaled by $1/c$ with respect to the electric 
one. We will come back later, in Section 2.2 and Section 4.1, to the 
consequences of this choice. 

%%%%%%%%%%%%%%%%%%%%%%%%%%%%%%%%%%%%%%%%%%%%%%%%%%%%%%%%%%%%%%%
\subsection{Conformal rescaling and FRW background subtraction}
%%%%%%%%%%%%%%%%%%%%%%%%%%%%%%%%%%%%%%%%%%%%%%%%%%%%%%%%%%%%%%%

With the purpose of studying gravitational instability in a FRW background, 
it is convenient to factor out the homogeneous and isotropic solutions of 
the above equations. To this aim we also perform a conformal rescaling
of the metric with conformal factor $a(t)$, the scale--factor of
FRW models, and change the time variable to the conformal time $\tau$, 
defined by $d\tau = dt / a(t)$. 

The line--element is then written in the form 
\be
ds^2 = a^2(\tau)\big[ - c^2 d\tau^2 + \gamma_{\alpha\beta}({\bf q}, \tau) 
dq^\alpha d q^\beta \big] \;, 
\ee
where $a^2(\tau) \gamma_{\alpha\beta}({\bf q}, \tau) \equiv 
h_{\alpha\beta}({\bf q}, t(\tau))$. 
For later convenience we fix the Lagrangian coordinates $q^\alpha$ to have 
physical dimension of length and the conformal time variable $\tau$ to have 
dimension of time. As a consequence the spatial metric $\gamma_{\alpha\beta}$
is dimensionless, as is the scale--factor $a(\tau)$ which must be 
determined by solving the Friedmann equations for a perfect fluid of dust 
\be
\biggl({a' \over a}\biggr)^2 = {8\pi G \over 3} \varrho_b a^2 - \kappa c^2 \;,
\ee
\be 
2 {a'' \over a} - \biggl({a' \over a}\biggr)^2 + \kappa c^2 = 0 \;. 
\ee
Here primes denote differentiation with respect to the conformal time $\tau$
and $\kappa$ represents the curvature parameter of FRW models, 
which, because of our choice of dimensions, cannot be normalized as usual. 
So, for an Einstein--de Sitter universe $\kappa=0$, but for a closed
(open) model we simply have $\kappa>0$ ($\kappa<0$). 
Let us also note that the curvature parameter is related to a 
Newtonian squared time--scale $\kappa_N$ through $\kappa_N \equiv \kappa c^2$
(e.g. Peebles 1980; Coles \& Lucchin 1995); in other words $\kappa$ is an 
intrinsically PN quantity. 

By subtracting the isotropic Hubble--flow, we introduce a {\em peculiar 
velocity--gradient tensor}
\be
\vartheta^\alpha_{~\beta} \equiv a ~c \tilde u^\alpha_{~;\beta} - {a' \over a} 
\delta^\alpha_{~\beta} = {1 \over 2} \gamma^{\alpha\gamma} 
{\gamma_{\gamma\beta}}' \;,
\ee
where $\tilde u^a = (1/a,0,0,0)$. 

Thanks to the introduction of this tensor we can rewrite the Einstein's 
equations in a more cosmologically convenient form. 
The energy constraint becomes 
\be
\vartheta^2 - \vartheta^\mu_{~\nu} \vartheta^\nu_{~\mu} + 4 {a' \over a} 
\vartheta + c^2 \bigl( {\cal R} - 6 \kappa \bigr) = 16 \pi G a^2 
\varrho_b \delta \;,
\ee
where ${\cal R}^\alpha_{~\beta}(\gamma) = a^2~^{(3)}\! R^\alpha_{~\beta}(h)$ 
is the conformal Ricci curvature of the three--space, i.e. that corresponding 
to the metric $\gamma_{\alpha\beta}$; for the background FRW solution 
$\gamma^{FRW}_{\alpha\beta} = (1 + {\kappa\over 4} q^2)^{-2} 
\delta_{\alpha\beta}$, one has ${\cal R}^\alpha_{~\beta}(\gamma^{FRW}) 
= 2 \kappa 
\delta^\alpha_{~\beta}$. We also introduced the density contrast 
$\delta \equiv (\varrho - \varrho_b) /\varrho_b$. 

The momentum constraint reads 
\be
\vartheta^\alpha_{~\beta||\alpha} = \vartheta_{,\beta} \;, 
\ee
where the double vertical bars denote covariant derivatives in the 
three--space with metric $\gamma_{\alpha\beta}$. 

Finally, after replacing the density from the energy constraint and
subtracting the background contribution, the extrinsic curvature evolution 
equation becomes 
\be
{\vartheta^\alpha_{~\beta}}' + 2 {a' \over a} \vartheta^\alpha_{~\beta} + 
\vartheta \vartheta^\alpha_{~\beta} + {1 \over 4} 
\biggl( \vartheta^\mu_{~\nu} \vartheta^\nu_{~\mu} - \vartheta^2 \biggr) 
\delta^\alpha_{~\beta} + {c^2 \over 4} \biggl[ 4 {\cal R}^\alpha_{~\beta} 
- \bigl( {\cal R} + 2 \kappa \bigr) \delta^\alpha_{~\beta} \biggr]
= 0 \;. 
\ee

The Raychaudhuri equation for the evolution of the 
{\em peculiar volume--expansion scalar} $\vartheta$ becomes 
\be
\vartheta' + {a' \over a} \vartheta + \vartheta^\mu_{~\nu} \vartheta^\nu_{~\mu} 
+ 4 \pi G a^2 \varrho_b \delta =0 \;. 
\ee
The main advantage of this formalism is that there is only one dimensionless 
(tensor) variable in the equations, namely the spatial metric tensor 
$\gamma_{\alpha\beta}$, which is present with its partial time derivatives 
through $\vartheta^\alpha_{~\beta}$ [Eq.(19) above], 
and with its spatial gradients through the spatial Ricci 
curvature ${\cal R}^\alpha_{~\beta}$. The only remaining variable is the 
density contrast which can be written in the form
\be
\delta({\bf q}, \tau) = (1 + \delta_0({\bf q})) \bigl[\gamma({\bf q}, \tau)/ 
\gamma_0 ({\bf q}) \bigr]^{-1/2} - 1 \;,
\ee
where $\gamma \equiv {\rm det} ~\gamma_{\alpha\beta}$. 
A relevant advantage of having a single tensorial variable, for our purposes, 
is that there can be no extra powers of $c$ hidden in the definition of 
different quantities. 

%%%%%%%%%%%%%%%%%%%%%%%%%%%%%%%%%%%%%%%%%%%%%%%%%%%%%%%%%%%%%%%
\subsection{Fluid--flow approach }
%%%%%%%%%%%%%%%%%%%%%%%%%%%%%%%%%%%%%%%%%%%%%%%%%%%%%%%%%%%%%%%

Following the {\em fluid--flow} approach, described in the classical review by 
Ellis (1971) [see also Ehlers (1993)], we can alternatively describe our system 
in terms of fluid properties, in our case matter density, 
volume--expansion scalar and shear tensor, and two geometric tensors, the 
electric and magnetic parts of the Weyl tensor defined above. The 
derivation of the equations reported below is thoroughly described by Ellis 
(1971) and will not be reported here. 

For most cosmological purposes it is convenient to adopt the conformal 
rescaling and FRW background subtraction described in the previous 
sub--section. 
Therefore, we can start by writing the continuity equation directly in terms 
of the density contrast $\delta$,
\be 
{D \delta \over D \tau} + \bigl(1 + \delta \bigr) \vartheta = 0 \;, 
\ee
with $D \over D \tau$ denoting convective differentiation with respect to the 
conformal time $\tau$. In our Lagrangian frame, however, and for a scalar 
field, convective differentiation and partial differentiation coincide. 
The formal solution of this equation is given by Eq.(24) above. 
The peculiar volume--expansion scalar $\vartheta$ obeys 
the Raychaudhuri equation 
which we can rewrite in the form 
\be 
{D \vartheta \over D \tau} + {a' \over a} \vartheta + {1 \over 3} \vartheta^2
+ \sigma^\alpha_{~\beta} \sigma^\beta_{~\alpha} + 
4 \pi G a^2 \varrho_b \delta = 0 \;,
\ee
where $\sigma^\alpha_{~\beta} \equiv \vartheta^\alpha_{~\beta} - 
{1 \over 3} \delta^\alpha_{~\beta} \vartheta$ 
is the shear tensor. The shear, in turn, evolves according to 
\be
{D \sigma^\alpha_{~\beta} \over D \tau} + {a' \over a} \sigma^\alpha_{~\beta} 
+ {2 \over 3} \vartheta \sigma^\alpha_{~\beta} + \sigma^\alpha_{~\gamma} 
\sigma^\gamma_{~\beta} - {1 \over 3} \delta^\alpha_{~\beta} 
\sigma^\gamma_{~\delta} \sigma^\delta_{~\gamma} + {\cal E}^\alpha_{~\beta} = 
0 \;,
\ee
where we have rescaled the electric tide as ${\cal E}^\alpha_{~\beta} \equiv 
a^2 E^\alpha_{~\beta}$, which can be written in terms of our
new variables as 
\be
{\cal E}^\alpha_{~\beta} = {1 \over 3} \delta^\alpha_{~\beta} 
\sigma^\mu_{~\nu} \sigma^\nu_{~\mu} + {1 \over 3} \vartheta 
\sigma^\alpha_{~\beta} + {a' \over a} \sigma^\alpha_{~\beta} - 
\sigma^\alpha_{~\gamma} \sigma^\gamma_{~\beta} 
+ c^2 \biggl( {\cal R}^\alpha_{~\beta} - {1 \over 3} \delta^\alpha_{~\beta} 
{\cal R} \biggr) \;.
\ee

Note that, for a generic second rank tensor $A^\alpha_{~\beta}$, one has 
\be
{D A^\alpha_{~\beta} \over D \tau} = {d A^\alpha_{~\beta} \over d \tau} + 
\sigma^\alpha_{~\gamma} A^\gamma_{~\beta} - \sigma^\gamma_{~\beta} 
A^\alpha_{~\gamma} \;,
\ee
where ${d \over d \tau}$ denotes the total derivative with respect to $\tau$,
which in comoving coordinates coincides with the partial one. 
The two last terms in the r.h.s. come from writing the 
Christoffel symbols in our gauge. It is then clear 
that when the ${D \over D \tau}$ operator acts on either the shear or 
the complete $\vartheta^\alpha_{~\beta}$ tensor, the second and third term 
in the r.h.s. cancel each other and the convective and total 
differentiation coincide. 
This cancellation also occurs
for a generic $A^\alpha_{~\beta}$ if either the relevant Christoffel symbols 
vanish (as it is the case for the Newtonian limit in Eulerian coordinates) or 
the convective derivative acts on the 
eigenvalues of $A^\alpha_{~\beta}$ and such a tensor has the same eigenvectors 
of $\sigma^\alpha_{~\beta}$ [as it is the case for the electric tide in the 
``silent universe" case (Barnes \& Rowlingson 1989; Matarrese, Pantano \& Saez 
1993; Bruni, Matarrese \& Pantano 1995b)]. 

The electric tidal tensor evolves according to 
\begin{eqnarray}
\nonumber 
{D {\cal E}^\alpha_{~\beta} \over D \tau} + {a' \over a} 
{\cal E}^\alpha_{~\beta} 
+ \vartheta {\cal E}^\alpha_{~\beta} + \delta^\alpha_{~\beta} 
\sigma^\gamma_{~\delta} {\cal E}^\delta_{~\gamma} - {3 \over 2} \biggl(
{\cal E}^\alpha_{~\gamma} \sigma^\gamma_{~\beta} + 
\sigma^\alpha_{~\gamma} {\cal E}^\gamma_{~\beta} \biggr) - 
\\ 
- {c^2 \over 2} \gamma_{\beta\eta} \biggl({\tilde \eta}^{\eta\gamma\delta} 
{\cal H}^\alpha_{~\gamma||\delta} + {\tilde \eta}^{\alpha\gamma\delta} 
{\cal H}^\eta_{~\gamma||\delta} \biggr) + 4\pi G a^2 \varrho_b(1+ \delta) 
\sigma^\alpha_{~\beta} = 0 \;,  
\end{eqnarray}
where we have rescaled the magnetic tide as ${\cal H}^\alpha_{~\beta} \equiv 
a^2 H^\alpha_{~\beta}$ and redefined the Levi--Civita tensor so that 
$\eta^{\alpha\beta\gamma} = \gamma^{-1/2} \epsilon^{\alpha\beta\gamma}$
(for simplicity we used the same symbol after rescaling). 

Finally, the magnetic Weyl tensor evolves according to
\begin{eqnarray}
\nonumber 
{D {\cal H}^\alpha_{~\beta} \over D \tau} + {a' \over a} 
{\cal H}^\alpha_{~\beta} 
+ \vartheta {\cal H}^\alpha_{~\beta} + \delta^\alpha_{~\beta} 
\sigma^\gamma_{~\delta} {\cal H}^\delta_{~\gamma} - {3 \over 2} \biggl(
{\cal H}^\alpha_{~\gamma} \sigma^\gamma_{~\beta} + 
\sigma^\alpha_{~\gamma} {\cal H}^\gamma_{~\beta} \biggr) + 
\\ 
+ {1 \over 2} \gamma_{\beta\eta} \biggl(\eta^{\eta\gamma\delta} 
{\cal E}^\alpha_{~\gamma||\delta} + \eta^{\alpha\gamma\delta} 
{\cal E}^\eta_{~\gamma||\delta} \biggr) = 0 \;. 
\end{eqnarray}
Note that, following the discussion above, in the last two equations 
the convective time derivative must include the two terms proportional to the 
shear, as in Eq.(29). Note that, apart from the cases listed after Eq.(29), 
these two terms cannot be disregarded even in the Newtonian limit. 

In the fluid--flow approach, besides the evolution equations, 
one has to satisfy a number of constraint equations. 
One has: the momentum constraint, which we rewrite in the form 
\be
\sigma^\alpha_{~\beta||\alpha} = {2 \over 3} \vartheta_{,\beta} \;, 
\ee
the ${\cal H}$--$\sigma$ constraint (which we actually used in Section 2 to 
define the magnetic tide in terms of derivatives of the spatial 
metric), 
\be
{\cal H}^\alpha_{~\beta} = {1 \over 2} \gamma_{\beta\mu} 
\biggl(\eta^{\mu\gamma\delta} 
\sigma^\alpha_{~\gamma||\delta} + \eta^{\alpha\gamma\delta} 
\sigma^\mu_{~\gamma||\delta} \biggr) \;, 
\ee
the ${\rm div} ~{\cal E}$ constraint, 
\be
{\cal E}^\alpha_{~\beta||\alpha} = - \gamma_{\beta\mu}\gamma_{\alpha\nu} 
\eta^{\mu\lambda\gamma}\sigma^\nu_{~\lambda} {\cal H}^\alpha_{~\gamma} 
+ {8 \pi G \over 3} a^2 \varrho_b \delta_{,\beta} \;,
\ee 
and the ${\rm div} ~{\cal H}$ constraint 
\be
c^2 {\cal H}^\alpha_{~\beta||\alpha} = \gamma_{\beta\mu}\gamma_{\alpha\nu} 
\eta^{\mu\lambda\gamma}\sigma^\nu_{~\lambda} {\cal E}^\alpha_{~\gamma} \;.
\ee

In the above equations one also needs to know the three--metric 
$\gamma_{\alpha\beta}$. This can be obtained from the 
evolution equation 
\be
{\gamma_{\alpha\beta}}' = 2 \gamma_{\alpha\gamma} \vartheta^\gamma_{~\beta} \;,
\ee
which is however only valid in our Lagrangian coordinates. 
In order to completely fix the spatial dependence of the metric one also 
needs to specify the energy constraint (the trace of the Gauss--Codacci 
relations), which we rewrite in the form
\be 
c^2 \bigl( {\cal R} - 6 \kappa \bigr) = \sigma^\alpha_{~\beta}
\sigma^\beta_{~\alpha} - {2 \over 3} \vartheta^2 - 4 {a' \over a}
\vartheta + 16 \pi G a^2 \varrho_b \delta \;.
\ee

Although we will not use the fluid--flow approach in this paper it
is interesting to have the complete form of the equations, with the correct 
powers of $c^2$ included, in order to understand the Newtonian meaning of 
the electric and magnetic tide. 
We will come back to this point in Section 4.1.

%%%%%%%%%%%%%%%%%%%%%%%%%%%%%%%%%%%%%%%%%%%%%%%%%%%%%%%%%%%%%%%
\subsection{Local Eulerian coordinates }
%%%%%%%%%%%%%%%%%%%%%%%%%%%%%%%%%%%%%%%%%%%%%%%%%%%%%%%%%%%%%%%

Our intuitive notion of Eulerian coordinates, involving a universal absolute 
time and globally flat spatial coordinates is intimately Newtonian; 
nevertheless it is possible to construct a local coordinates system 
which reproduces this picture for a suitable set of observers. 
This issue has been already addressed by Matarrese, Pantano \& Saez 
(1994a,b), who introduced local Eulerian -- FRW comoving -- coordinates $x^A$ 
which are related to the Lagrangian ones $q^\alpha$ via the Jacobian matrix 
with elements
\be
{\cal J}^A_{~~\alpha} ({\bf q},\tau) \equiv 
{\partial x^A \over \partial q^\alpha} \equiv 
\delta^A_{~\alpha} + {\cal D}^A_{~\alpha} ({\bf q},\tau) \;, 
\ \ \ \ \ \ A=1,2,3 \;,
\ee
where ${\cal D}^A_{~\alpha} ({\bf q},\tau)$ is called
{\em deformation tensor}. Each matrix element ${\cal J}^A_{~~\alpha}$
labelled by the Eulerian index $A$ can 
be thought as a three--vector, namely a {\em triad}, defined on the 
hypersurfaces of constant conformal time. 
As shown in (Matarrese, Pantano \& Saez 
1994a,b), they evolve according to 
\be
{{\cal J}^A_{~~\alpha}}' = \vartheta^\gamma_{~\alpha} {\cal J}^A_{~~\gamma} \;,
\ee
which also follows from the condition of parallel transport of the triads 
relative to ${\bf q}$ along the world--line of the corresponding 
fluid element $D(a {{\cal J}^A_{~~\alpha}}) / D t =0$ (see also Kasai 1995). 

Our local Eulerian coordinates are such that the 
spatial metric takes the Euclidean form $\delta_{AB}$,
i.e. 
\be
\gamma_{\alpha\beta} ({\bf q},\tau) = \delta_{AB} {\cal J}^A_{~~\alpha} 
({\bf q},\tau) {\cal J}^B_{~~\beta} ({\bf q},\tau) \;. 
\ee
Correspondingly the matter density can be rewritten in the suggestive form
\be
\varrho ({\bf q},\tau) = \varrho_b(\tau) \bigl( 1 + \delta_0({\bf q}) \bigr)
\bigl[ {\cal J}({\bf q},\tau) / {\cal J}_0 ({\bf q}) \bigr]^{-1} \;,
\ee
where ${\cal J} \equiv {\rm det} {\cal J}^A_{~~\alpha}$. Note that, contrary to
the Newtonian case, it is generally impossible in GR to fix ${\cal J}_0=1$,
as this would imply that the initial Lagrangian space is conformally flat, 
which is only possible if the initial perturbations vanish. 

%%%%%%%%%%%%%%%%%%%%%%%%%%%%%%%%%%%%%%%%%%%%%%%%%%%%%%%%%%%%%%%%%%%%%%%
\subsection{Linear perturbation theory in Lagrangian coordinates}
%%%%%%%%%%%%%%%%%%%%%%%%%%%%%%%%%%%%%%%%%%%%%%%%%%%%%%%%%%%%%%%%%%%%%%%

In this subsection we will deal with the linearization of the equations 
obtained in Section 2.1. This will be done mostly for pedagogical purposes, in 
that it will allow us to obtain a number of results which will turn out 
to be useful for the $1/c^2$ expansion. Apart from this, it can be interesting 
to re--obtain the classical results of linear theory in the comoving and 
synchronous gauge only in terms of the spatial metric coefficients. 

Let us then write the spatial metric tensor of the physical 
(i.e. perturbed) space--time in the form 
\be
\gamma_{\alpha\beta} = {\bar \gamma}_{\alpha\beta} + w_{\alpha\beta} \;,
\ee
with ${\bar \gamma}_{\alpha\beta}$ the spatial metric of the 
background space -- in our case the maximally symmetric FRW one,
${\bar \gamma}_{\alpha\beta} = \gamma^{FRW}_{\alpha\beta}$ -- and 
$w_{\alpha\beta}$ a small perturbation. Also, we assume that the only
non--geometric quantity in our equations, namely the initial density contrast
$\delta_0$, is everywhere much smaller than unity. 

As usual, we can take advantage of the maximal symmetry of the background FRW 
spatial sections to classify metric perturbations as scalars, vectors and 
tensors (e.g. Bardeen 1980). We then write
\be
w_{\alpha\beta} = \chi {\bar \gamma}_{\alpha\beta} + \zeta_{|\alpha\beta} +
{1 \over 2} \bigl( \xi_{\alpha | \beta} + \xi_{\beta | \alpha} \bigr) 
+ \pi_{\alpha\beta} \;, 
\ee
with
\be
\xi^\alpha_{~~|\alpha}= \pi^\alpha_{~\alpha} = \pi^\alpha_{~\beta|\alpha}=0 \;,
\ee
where a single vertical bar is used for covariant differentiation in the 
background three--space with metric ${\bar \gamma}_{\alpha\beta}$. 
In the above decomposition $\chi$ and $\zeta$ represent scalar modes, 
$\xi^\alpha$ vector modes and $\pi^\alpha_{~\beta}$ tensor modes 
(indices being raised by the contravariant background three--metric). 

Before entering into the discussion of the equations for these perturbation 
modes, let us quote a result which will be also useful in the next sections. 
In the $\vartheta^\alpha_{~\beta}$ evolution equation and in the energy 
constraint the combination ${\cal P}^\alpha_{~\beta} \equiv 
4 {\cal R}^\alpha_{~\beta} - \bigl( {\cal R} + 
2 \kappa \bigr) \delta^\alpha_{~\beta}$ and its trace appear. 
To first order in the metric perturbation one has
\be
{\cal P}^\alpha_{~\beta} (w) = -2 \biggl[ \bigl(\nabla^2 - 2 \kappa \bigr) 
\pi^\alpha_{~\beta} + {\chi_|}^\alpha_{~\beta} + \kappa \chi 
\delta^\alpha_{~\beta} \biggr] \;,
\ee
where $\nabla^2 ( \cdot ) \equiv {( \cdot )_|}^\gamma_{~\gamma}$. 
Only the scalar mode $\chi$ and the tensor modes contribute to the 
three--dimensional Ricci curvature. 

As well known, in linear theory scalar, vector and tensor modes are 
independent. The equation of motion for the tensor modes 
is obtained by linearizing the traceless part of the 
$\vartheta^\alpha_{~\beta}$ evolution equation, Eq.(22). One has
\be 
{\pi_{\alpha\beta}}'' + 2 {a' \over a} {\pi_{\alpha\beta}}' 
- c^2 \bigl(\nabla^2 - 2 \kappa \bigr) \pi_{\alpha\beta} = 0 \;,
\ee
which is the equation for the free propagation of gravitational 
waves in a FRW background (compare with Eq.(3) in the Einstein--de Sitter 
case). The general solution of this equation is well known (e.g. Weinberg 
1972) and will not be reported here. 

At the linear level, in the irrotational case, the two vector modes represent 
gauge modes which can be set to zero, $\xi^\alpha=0$. 

The two scalar modes are linked together through the momentum constraint, 
which leads to the relation
\be
\chi = \chi_0 + \kappa (\zeta - \zeta_0) \;.
\ee
The energy constraint gives
\be
\bigl( \nabla^2 + 3 \kappa \bigr) \biggl[ {a' \over a} \zeta' + 
\bigl(4 \pi G a^2 \varrho_b - \kappa c^2 \bigr) 
(\zeta - \zeta_0) - c^2 \chi_0 \biggr] = 8 \pi G a^2 \varrho_b \delta_0 \;, 
\ee
while the evolution equation gives
\be
\zeta'' + 2 {a' \over a} \zeta' = c^2 \chi \;.
\ee

An evolution equation only for the scalar mode $\zeta$ can be obtained 
by combining together the evolution equation and the energy constraint;
it reads 
\be
\bigl( \nabla^2 + 3 \kappa \bigr) \biggl[ \zeta'' + {a' \over a} \zeta' 
- 4\pi G a^2 \varrho_b (\zeta - \zeta_0) \biggr] = - 8 \pi G a^2 
\varrho_b \delta_0 \;. 
\ee

On the other hand, linearizing the solution of the continuity 
equation, Eq.(24), gives 
\be
\delta = \delta_0 - {1 \over 2} (\nabla^2 + 3 \kappa ) (\zeta - \zeta_0) \;,
\ee
which replaced in the previous equation gives 
\be
\delta'' + {a' \over a} \delta' - 4\pi G a^2 \varrho_b \delta = 0 \;.
\ee
This is the well--known equation for linear density fluctuations, whose 
general solution can be found in (Peebles 1980). Once $\delta(\tau)$ is known, 
one can easily obtain $\zeta$ and $\chi$, which completely solves the linear 
problem. 

Eq.(50) above has been obtained in whole generality; we could have used 
instead the well--known residual gauge ambiguity of the synchronous 
coordinates to simplify its form. In fact, $\zeta$ is determined up to a 
space--dependent scalar, which would neither contribute to the spatial 
curvature, nor to the velocity--gradient tensor. For instance, we 
could fix $\zeta_0$ so that $(\nabla^2 + 3 \kappa) \zeta_0 = - 2 \delta_0$,
so that the $\zeta$ evolution equation takes the same form as that for 
$\delta$. 

In order to better understand the physical meaning of the two 
scalar modes $\chi$ and $\zeta$, let us consider the simplest case of an 
Einstein--de Sitter background ($\kappa=0$), for which 
$a(\tau) \propto \tau^2$. By fixing the gauge so that 
$\nabla^2 \zeta_0 = - 2 \delta_0$ we obtain $\chi(\tau)=\chi_0$ and 
\be 
\zeta(\tau) = {c^2 \over 10} \chi_0 \tau^2 + B_0 \tau^{-3} \;,
\ee
where the amplitude $B_0$ of the decaying mode is an arbitrary function of the 
spatial coordinates. Consistency with the Newtonian limit suggests 
$\chi_0 \equiv - {10 \over 3c^2} \varphi_0$, with $\varphi_0$ the initial 
peculiar gravitational potential, related to $\delta_0$ through 
$\nabla^2 \varphi_0 = 4 \pi G a_0^2 \varrho_{0b} \delta_0$. We can then write 
\be
\zeta(\tau) = - {1 \over 3} \varphi_0 \tau^2 + B_0 \tau^{-3} \;.
\ee

This result clearly shows that, at the Newtonian level, the linearized 
metric is 
\be
\gamma_{\alpha\beta} = \delta_{\alpha\beta} + \zeta_{|\alpha\beta} \;,
\ee
while the perturbation mode $\chi$ is already 
PN. Note that also the tensor modes are at least PN. 

These results also confirm the above conclusion that in the general GR case 
the initial Lagrangian spatial metric cannot be flat, i.e. 
${\cal J}_0 \neq 1$, because of the initial ``seed" PN metric perturbation 
$\chi_0$.

%%%%%%%%%%%%%%%%%%%%%%%%%%%%%%%%%%%%%%%%%%%%%%%%%%%%%%%%%%%%%%%%%%%%%%%
\section{Newtonian approximation } 
%%%%%%%%%%%%%%%%%%%%%%%%%%%%%%%%%%%%%%%%%%%%%%%%%%%%%%%%%%%%%%%%%%%%%%%

The Newtonian equations in Lagrangian form can be obtained from the 
full GR equations of Section 2.1 by an expansion in inverse 
powers of the speed of light; as a consequence of our 
gauge choice, however, no odd powers of $c$ appear in the 
equations, which implies that the expansion parameter can be 
taken to be $1/c^2$. The physical meaning of this expansion has been 
already outlined in Section 1. 

Let us then expand the spatial metric in a form analogous to that 
used in our linear perturbation analysis of Section 2.4: 
\be
\gamma_{\alpha\beta} = {\bar \gamma}_{\alpha\beta} + {1 \over c^2} 
w^{(PN)}_{\alpha\beta} + {\cal O} \biggl( {1 \over c^4} \biggr) \;,
\ee
where we made explicit the $c$ dependence of the metric perturbation. 
The actual convergence of the series requires that the 
PN metric perturbation $\frac{1}{c^2} w^{(PN)}_{\alpha\beta}$ is much 
smaller than the background Newtonian metric 
${\bar \gamma}_{\alpha\beta}$. 
Let us first concentrate on the Newtonian metric; the properties of 
$w_{\alpha\beta}$ will be instead considered in Section 4. 

To lowest order in our expansion, the extrinsic curvature evolution equation,
Eq.(22), and the energy constraint, Eq.(20), imply that 
${\bar {\cal P}}^\alpha_{~\beta} \equiv 
{\cal P}^\alpha_{~\beta} ({\bar \gamma}) =0$, and recalling that 
$\kappa=\kappa_N/c^2$, 
\be 
{\bar {\cal R}}^\alpha_{~\beta} \equiv {\cal R}^\alpha_{~\beta} ({\bar \gamma}) 
= 0 \;:
\ee 
{\em in the Newtonian limit the spatial curvature identically vanishes}
(e.g. Ellis 1971). This important conclusion implies that 
${\bar \gamma}_{\alpha\beta}$ can be transformed to $\delta_{AB}$
globally, i.e. that one can write 
\be
{\bar \gamma}_{\alpha\beta} = \delta_{AB} {\bar {\cal J}}^A_{~~\alpha} 
{\bar {\cal J}}^B_{~~\beta} \;,
\ee
with integrable Jacobian matrix coefficients. In other words, at each time 
$\tau$ there exist {\em global Eulerian coordinates} $x^A$ such that 
\be
{\bf x}({\bf q}, \tau) = {\bf q} + {\bf S}({\bf q}, \tau) \;,
\ee
where ${\bf S}({\bf q}, \tau)$ is called the {\em displacement vector}, 
and the deformation tensor becomes in this limit 
\be
{\bar {\cal D}}^A_{~\alpha} = {\partial S^A \over \partial q^\alpha} \;.
\ee

The Newtonian Lagrangian metric can therefore be written in the form 
\be
{\bar \gamma}_{\alpha\beta}({\bf q}, \tau) = \delta_{AB} 
\biggl(\delta^A_{~\alpha} + {\partial S^A({\bf q}, \tau) 
 \over \partial q^\alpha} \biggr) 
\biggl(\delta^B_{~\beta} + {\partial S^B({\bf q}, \tau) 
\over \partial q^\beta} \biggr) \;.
\ee 

We can rephrase the above result as follows: the Lagrangian spatial metric in 
the Newtonian limit is that of Euclidean three--space in time--dependent 
curvilinear coordinates $q^\alpha$, defined at each time $\tau$ in terms of 
the Eulerian ones $x^A$ by inverting Eq.(59) above. As a consequence, the 
Christoffel symbols involved in spatial covariant derivatives (which we will 
indicate by a single bar or by a nabla operator followed by greek indices) 
do not vanish, but the vanishing of the spatial curvature implies that 
these covariant derivatives always commute. 

Contrary to the evolution equation and the energy constraint, the 
Raychaudhuri equation, Eq.(23) and the momentum constraint, Eq.(21), 
contain no explicit powers of $c$, and therefore preserve their form in going 
to the Newtonian limit. These equations therefore determine the background 
Newtonian metric ${\bar \gamma}_{\alpha\beta}$, i.e. they govern the evolution 
of the displacement vector ${\bf S}$. 

The Raychaudhuri equation becomes the master equation for the Newtonian 
evolution; it takes the form 
\be 
{\bar \vartheta}' + {a' \over a} {\bar \vartheta} + 
{\bar \vartheta}^\mu_{~\nu} {\bar \vartheta}^\nu_{~\mu} + 
4 \pi G a^2 \varrho_b \bigl( {\bar \gamma}^{-1/2} - 1 \bigr) = 0 \;,
\ee
where
\be
{\bar \vartheta}^\alpha_{~\beta} \equiv {1 \over 2} 
{\bar \gamma}^{\alpha\gamma} {\bar \gamma_{\gamma\beta}}' \;, 
\ee
and, for simplicity, we assumed $\delta_0=0$ (a restriction which is, however,
not at all mandatory). We also used the residual gauge freedom 
of our coordinate system to set ${\bar \gamma}_{\alpha\beta}(\tau_0) = 
\delta_{\alpha\beta}$, implying ${\bar {\cal J}}_0=1$, i.e. to make 
Lagrangian and Eulerian coordinates coincide at the initial time. 
That this choice is indeed possible in the Newtonian limit can be understood 
from our previous linear analysis, where this is achieved by taking, e.g., 
$\zeta_0=0$. 

The momentum constraint, 
\be
{\bar \vartheta}^\mu_{~\nu |\mu } = {\bar \vartheta}_{,\nu} \;, 
\ee
is actually related to the irrotationality assumption. 
We will come back to this point in the next section. 

Before closing this section, let us notice a general property of our 
expression for the Lagrangian metric: at each time $\tau$ it can be 
diagonalized by going to the local and instantaneous principal axes
of the deformation tensor. Calling $\bar \gamma_\alpha$ 
the eigenvalues of the 
metric tensor, ${\bar {\cal J}}_\alpha$ those of the Jacobian and 
${\bar d}_\alpha$ those of the deformation tensor, 
one has
\be
\bar \gamma_\alpha ({\bf q},\tau) = {\bar {\cal J}}_\alpha^2({\bf q},\tau) = 
\bigl[ 1 + {\bar d}_\alpha ({\bf q},\tau) \bigr]^2 \;.
\ee

In Section 3.2 below, the diagonal form of the metric tensor will be 
reconsidered in the frame of the Zel'dovich approximation. Beyond the 
mildly non--linear regime, where this approximation is consistently applied, 
diagonalizing the metric is in general, i.e. apart from 
specific initial configurations, of smaller practical use, because 
metric (and deformation) tensor, shear and tide generally have different 
eigenvectors. 

From this expression it becomes evident that, at shell--crossing, 
where some of the Jacobian eigenvalues go to zero, the related covariant 
metric eigenvalues just vanish. On the other hand, other quantities, like 
the matter density, the peculiar volume expansion scalar and some eigenvalues 
of the shear and tidal tensor will generally diverge at the location of the 
caustics (see Bruni, Matarrese \& Pantano 1995b, for a discussion). 
This diverging behaviour makes the description of the system 
extremely involved after this event. Although dealing with this problem is 
far outside the aim of the present paper, let us just mention that a number of 
ways out are available. One can convolve the various dynamical variables by a 
suitable low--pass filter, either at the initial time, in order to postpone 
the occurrence of shell--crossing singularities (e.g. Coles, Melott \& 
Shandarin 1993; Kofman et al. 1994), or at the time when they 
form, in order to smooth out the singular behaviour (e.g. Nusser \& Dekel 1992, 
and references therein); alternatively one 
can abandon the perfect fluid picture and resort to a discrete point--like 
particle set, which automatically eliminates the possible occurrence of 
caustics, at least for generic initial data. 
At this level, anyway, we prefer to take a conservative 
point of view and assume that the  actual range of validity of our formalism 
is up to shell--crossing. 

%%%%%%%%%%%%%%%%%%%%%%%%%%%%%%%%%%%%%%%%%%%%%%%%%%%%%%%%%%%%%%%%%%%%%%%
\subsection{Jacobian approach }
%%%%%%%%%%%%%%%%%%%%%%%%%%%%%%%%%%%%%%%%%%%%%%%%%%%%%%%%%%%%%%%%%%%%%%%

A more direct way to deal with the Lagrangian Newtonian equations is
to write them in terms of the Jacobian matrix ${\cal J}^A_{~~\alpha}$.
This approach is obviously related to the more usual ones in terms of 
the displacement vector ${\bf S}$ or in terms of the deformation tensor 
${\cal D}^A_{~~\alpha}$ (Buchert 1989; Moutarde et al. 1991; Bouchet 
et al. 1992; Buchert 1992; Catelan 1995). The evolution equation 
has been explicitly written directly in terms of the Jacobian matrix by Buchert 
\& G\"otz (1987), Lachi\`eze--Rey (1993) and Catelan (1995). 

In order to rewrite the Raychaudhuri equation in terms of the Jacobian matrix, 
we notice that 
\be
{\vartheta^\alpha_{~\beta}}' = {\cal J}^\alpha_{~~A} {\cal J}^A_{~~\beta}~''
- \vartheta^\alpha_{~\mu} \vartheta^\mu_{~\beta} \;,
\ee
where we have introduced the inverse Jacobian matrix
\be
{\cal J}^\alpha_{~~A} \equiv {\partial q^\alpha \over \partial x^A} \;, 
\ee
where Eulerian indices are raised and lowered by the Kronecker symbol. 
To make explicit our notation, we just stress that elements of 
${\partial x^A \over \partial q^\alpha}$ will be characterized 
by a greek (i.e. Lagrangian) index subscript, while elements of the 
inverse matrix ${\partial q^\alpha \over \partial x^A}$ will 
be characterized by a greek index superscript. 

Replacing the latter identity into the Newtonian expression Eq.(62) yields
\be
{\bar {\cal J}}^\alpha_{~A} \bar {{\cal J}^A_{~~\alpha}}'' + 
{a' \over a} {\bar {\cal J}}^{-1} {\bar {\cal J}}' = 
4 \pi G a^2 \varrho_b \bigl( 1 - {\bar {\cal J}}^{-1} \bigr) \;. 
\ee

Note that this expression is, apart from the use of a different time variable,
identical to Eq.(60), in Catelan (1995) [see also Appendix A in (Buchert 
1989), and (Buchert 1992)]. 

We also notice that the parallel transport condition, Eq.(39), can
be rewritten in the form 
\be
\vartheta^\alpha_{~\beta} = {\cal J}^\alpha_{~A} {{\cal J}^A_{~~\beta}}' \;.
\ee

This equation, together with $\vartheta^\alpha_{~\beta} = \frac{1}{2} 
\gamma^{\alpha\gamma} {\gamma_{\gamma\beta}}'$ gives the general relation 
\be
{{\cal J}_{A\alpha}}' {\cal J}^A_{~~\beta} = {\cal J}_{A\alpha} 
{{\cal J}^A_{~~\beta}}' \;.
\ee
Replacing these relations in the momentum constraint we obtain 
in whole generality 
\be
{\cal J}^\alpha_{~~A} {{{\cal J}^A_{~~\beta}}'}_{||\alpha} +
{{\cal J}^\alpha_{~~A}}_{||\alpha} {{\cal J}^A_{~~\beta}}' = 
\bigl({\cal J}^{-1} {\cal J}' \bigr)_{,\beta} \;.
\ee

On the other hand, in the Newtonian limit we have 
\be
{\bar {\cal J}}^A_{~~\alpha,\beta} = {\bar {\cal J}}^A_{~~\beta,\alpha} \;, 
\ee
as it follows from the fact that $S^A_{~~,\alpha\beta} 
= S^A_{~~,\beta\alpha}$. Using this commutation property it is easy to 
verify that 
\be
{\bar \Gamma}^\alpha_{\beta\gamma} = {\bar {\cal J}}^\alpha_{~A} 
{\bar {\cal J}}^A_{~~\beta,\gamma} \;.
\ee 
Thanks to the latter relation and to the well--known matrix identity 
${\rm Tr} \ln {\bf J} = \ln {\rm det} {\bf J}$, it is straightforward to 
verify that the momentum constraint in the Newtonian limit becomes an 
identity. 
It is then clear that Eq.(70) is more fundamental than the momentum constraint:
it plays the role of an {\em irrotationality condition} written in Lagrangian 
space. This is of course equivalent to the standard form [compare with 
Eq.(59) in (Catelan 1995)] 
\be
\epsilon^{\alpha\beta\gamma} {\bar {\cal J}}^A_{~~\beta} 
{{\bar {\cal J}}_{A\gamma}}' = 0 \;.
\ee
This equation, together with the Raychaudhuri equation above, Eq.(68), 
completely determines the Newtonian problem, in terms of either the Jacobian 
matrix, the deformation tensor or the displacement vector. 

The very fact that we have been able to recover the standard equations 
for the Newtonian approximation in the Lagrangian picture, by starting 
from the Lagrangian GR treatment and expanding in powers of $1/c^2$, should
be considered as a further confirmation of the validity of our method. 

%%%%%%%%%%%%%%%%%%%%%%%%%%%%%%%%%%%%%%%%%%%%%%%%%%%%%%%%%%%%%%%%%%%%%%%
\subsection{Zel'dovich approximation } 
%%%%%%%%%%%%%%%%%%%%%%%%%%%%%%%%%%%%%%%%%%%%%%%%%%%%%%%%%%%%%%%%%%%%%%%

Having shown the equivalence of our method, in the Newtonian limit, with 
the standard one, it is now trivial to recover the Zel'dovich approximation 
(Zel'dovich 1970). This is obtained by expanding Eq.(68) and Eq.(70) to first 
order in the displacement vector. The result is 
\be
{\bf x} ({\bf q},\tau) = {\bf q} + D(\tau) \nabla \Phi_0({\bf q}) \;,
\ee
where only the growing mode solution $D(\tau)$ of Eq.(52) has been considered, 
and we introduced the potential $\Phi_0({\bf q})$, such that 
$\nabla_q^2 \Phi_0 = - \delta_0/D_0$, where $\nabla_q^2$ is the standard 
(i.e. Euclidean) Laplacian in Lagrangian coordinates; more in general, 
at this perturbative order covariant and partial derivatives with respect
to the $q^\alpha$ coincide. The potential $\Phi_0$ is easily related 
to the initial peculiar gravitational potential defined in Section 1, 
$\Phi_0 = - (4\pi G a_0^2 \varrho_{0b} D_0)^{-1} \varphi_0$. 

More interesting is to derive from the above expression the 
corresponding {\em Zel'dovich metric}. It reads 
\be
\gamma^{ZEL}_{\alpha\beta} ({\bf q},\tau) = \delta_{\gamma\delta} 
\biggl(\delta^\gamma_{~\alpha} + D(\tau) {\Phi_0,}^\gamma_{~\alpha} 
({\bf q}) \biggr) 
\biggl(\delta^\delta_{~\beta} + D(\tau) {\Phi_0,}^\delta_{~\beta} 
({\bf q}) \biggr) \;.
\ee

One can of course diagonalize this expression by going to the principal axes 
of the deformation tensor. Calling $\lambda_\alpha$ 
the eigenvalues of the matrix ${\Phi_0,}^\alpha_{~\beta}$, one finds 
\be
\gamma^{ZEL}_\alpha ({\bf q}, \tau) = \bigl[ 1 + D(\tau) 
\lambda_\alpha({\bf q})\bigr]^2  \;.
\ee

Note that, contrary to what has been commonly done so far in the literature, 
the metric tensor must be evaluated at second order in the displacement vector, 
in order to obtain back the correct Zel'dovich expressions for the dynamical 
variables (density, shear, etc ...). 

The above diagonal form of the metric allows a straightforward calculation of 
all the relevant quantities. The well--known expression for the mass density 
is consistently recovered, 
\be
\varrho^{ZEL} = \varrho_b \prod_\alpha
\bigl(1 + D \lambda_\alpha \bigr)^{-1} \;.
\ee
The peculiar velocity--gradient tensor has the same eigenframe of 
the metric; its eigenvalues read 
\be
\vartheta^{ZEL}_\alpha = {D' \lambda_\alpha 
\over 1 + D \lambda_\alpha } \;.
\ee
By summing over $\alpha$ the latter expression we can obtain the 
peculiar volume--expansion scalar 
\be
\vartheta^{ZEL} = \sum_\alpha 
{D' \lambda_\alpha \over 1 + D \lambda_\alpha} 
\ee
and then the shear eigenvalues 
\be 
\sigma^{ZEL}_\alpha = {D' \lambda_\alpha \over 1 + D \lambda_\alpha} -
{1 \over 3} \sum_\alpha 
{D' \lambda_\alpha \over 1 + D \lambda_\alpha} \;.
\ee
The electric tide comes out just proportional to the shear. 
Its eigenvalues read 
\be
{\cal E}^{ZEL}_\alpha = - 4 \pi G a^2 \varrho_b {D \over D'} 
~\sigma^{ZEL}_\alpha \;. 
\ee

These expressions for the shear and the tide completely agree with those 
obtained by Kofman \& Pogosyan (1995) and Hui \& Bertschinger (1995). 
The fact that metric, shear and tide have simultaneous eigenvectors shows that 
fluid elements in the Zel'dovich approximation actually evolve as in 
a ``silent universe" (Matarrese, Pantano \& Saez 1994a; Bruni Matarrese \& 
Pantano 1995b), with no influence from the environment, except 
for that implicit in the self--consistency of the initial conditions. 

So far the Zel'dovich approximation has been obtained by first 
taking the Newtonian limit ($c \to \infty$) of the GR equations and then 
linearizing them with respect to the Newtonian displacement vector. 
One could also drop the first step and linearize the GR equations of
Section 2.1 with respect to the local deformation tensor as 
introduced in Section 2.3; in such a case 
one would get a fully relativistic version of the Zel'dovich approximation. 

The latter problem has been already discussed a number of times 
by various authors. Unfortunately, there has been a lot of 
misunderstanding on what the ``relativistic Zel'dovich approximation" should 
actually be. Most authors just deal with the GR version of the Zel'dovich {\em 
solution}, i.e. with the non--linear evolution of planar perturbations, which 
is a sub--case of the well--known exact solutions obtained by Szekeres (1975). 
Such an approach, however, does not allow to deal with the approximate 
non--linear behaviour of {\em generic} perturbations in a relativistic 
framework. 

%%%%%%%%%%%%%%%%%%%%%%%%%%%%%%%%%%%%%%%%%%%%%%%%%%%%%%%%%%%%%%%%%%%%%%%
\subsection{Lagrangian Bernoulli equation }
%%%%%%%%%%%%%%%%%%%%%%%%%%%%%%%%%%%%%%%%%%%%%%%%%%%%%%%%%%%%%%%%%%%%%%%

As we have demonstrated above, it is always 
possible, in the frame of the Newtonian approximation, to define a global 
Eulerian picture. This will be the picture of the fluid evolution as given 
by an observer that, at the point ${\bf x} = {\bf q} + {\bf S}({\bf q}, \tau)$ 
and at the time $\tau$ observes the fluid moving with physical peculiar 
three--velocity ${\bf v} = d {\bf S}/d \tau$. From the point
of view of a Lagrangian observer, who is comoving with the fluid, 
the Eulerian observer, which is located at constant ${\bf x}$, is moving with 
three--velocity $d {\bf q} ({\bf x}, \tau)/d \tau = - {\bf v}$. 

The line--element characterizing the Newtonian approximation in the 
Eulerian frame is well known (e.g. Peebles 1980) 
\be
ds^2 = a^2(\tau) \biggl[ - \biggl(1 + {2\varphi_g ({\bf x}, \tau) \over c^2} 
\biggr) ~c^2 d\tau^2 + \delta_{AB} dx^A dx^B \biggr] \;, 
\ee
with $\varphi_g$ the peculiar gravitational potential, determined 
by the mass distribution through the Eulerian Poisson equation, 
\be
\nabla_x^2 \varphi_g ({\bf x}, \tau) = 4 
\pi G a^2(\tau) \varrho_b(\tau) \delta({\bf x}, \tau) \;,
\ee
where the Laplacian $\nabla_x^2$, as well as the nabla operator $\nabla$, 
have their standard Euclidean meaning. 
The perturbation in the time--time component of the metric tensor 
here comes from the different proper time of the Eulerian 
and Lagrangian observers. As already noticed in the Introduction, 
this Newtonian line--element is different to that of Eq.(2), describing the 
so--called weak--field limit; the extra term $- 2\varphi_g/c^2$, 
multiplying the spatial line--element, would in fact give rise 
to PN corrections in the equations of motion for the matter. 

It is now crucial to realize that all the dynamical equations obtained 
so far, being entirely expressed in terms of three--tensors, keep their form 
in going to the Eulerian picture, only provided the convective time 
derivatives of tensors of any rank (scalars, vectors and 
tensors) are modified as follows: 
\be
{D \over D \tau} \rightarrow {\partial \over \partial \tau} + 
{\bf v} \cdot \nabla \;, \ \ \ \ \ \ \ 
{\bf v} \equiv {d {\bf S} \over d \tau} \;.
\ee

This follows from the fact that, for the metric above, $\bar \Gamma^0_{AB} = 
\bar \Gamma^A_{0B} =\bar \Gamma^A_{BC}=0$, which also obviously implies that 
covariant derivatives with respect to $x^A$ reduce to partial ones. 

The irrotationality assumption now has the obvious consequence that we can 
define an Eulerian velocity potential $\Phi_v$ through
\be
{\bf v} ({\bf x}, \tau) = \nabla \Phi_v ({\bf x}, \tau) \;.
\ee
The Newtonian peculiar velocity--gradient tensor then becomes 
\be
\bar \vartheta_{AB} = {\partial^2 \Phi_v \over \partial x^A \partial x^B} \;,
\ee
because of which the momentum constraint gets trivially satisfied and the 
magnetic Weyl tensor becomes identically zero in the Newtonian limit. 

We can now write the Raychaudhuri equation for the Eulerian peculiar 
volume--expansion scalar $\bar \vartheta$, 
and use the Poisson equation to get, as a first spatial integral, the 
{\em Euler equation} 
\be
{\bf v}~' + {\bf v} \cdot \nabla {\bf v} + 
{a' \over a} {\bf v} = - \nabla \varphi_g \;.
\ee

This can be further integrated to give the {\em Bernoulli equation} 
\be
\Phi_v' + {a' \over a} \Phi_v + {1\over 2} \bigl(\nabla \Phi_v \bigr)^2 
= - \varphi_g \;.
\ee

On the other hand, by taking gradients of the Euler equation we can 
obtain an Eulerian evolution equation for the tensor $\bar \vartheta_{AB}$.
More interesting is that this equation can be transported back to the 
Lagrangian frame to get 
\be
{\bar \vartheta}^{\alpha~'}_{~\beta}
+ {a' \over a} \bar \vartheta^\alpha_{~\beta}  + 
\bar \vartheta^\alpha_{~\gamma}  
\bar \vartheta^\gamma_{~\beta}  = - {\varphi_{g|}^{(L)}}^\alpha_{~\beta} \;. 
\ee
where $\varphi_g^{(L)}$ must be thought as a {\em Lagrangian peculiar 
gravitational potential} to be 
determined through the {\em Lagrangian Poisson equation} 
\be
{\varphi_{g|}^{(L)}}^\alpha_{~\alpha} = 4 \pi G a^2 \varrho_b 
(\bar \gamma^{-1/2} -1) \;.
\ee

These two Lagrangian expressions will turn out to be very useful for the
PN calculations of Section 4. 

There is, however, another consequence of these equations that we can easily 
derive. While in the Lagrangian frame the three--velocity field does 
not exist, the tensor $\bar \vartheta^\alpha_{~\beta}$ is well--defined, 
so that we can rewrite Eq.(87) in Lagrangian coordinates to obtain a
{\em Lagrangian velocity potential} $\Phi_v^{(L)}({\bf q}, \tau)$,
through 
\be
\bar \vartheta^\alpha_{~\beta} \equiv {\Phi_{v|}^{(L)}}^\alpha_{~\beta} \;.
\ee
This potential obeys what we can name the {\em Lagrangian Bernoulli equation}, 
which is easily obtained from the Bernoulli equation above, provided we 
recollect the convective time derivative and express it in Lagrangian form. 
We get
\be 
{\Phi_v^{(L)}}' + {a' \over a} \Phi_v^{(L)} - {1\over 2} \gamma^{\alpha\beta} 
\Phi_{v,\alpha}^{(L)} \Phi_{v,\beta}^{(L)} = - \varphi_g^{(L)} \;.
\ee
The most astonishing difference between the Eulerian and Lagrangian 
versions of the Bernoulli equation is the relative sign of the temporal and
spatial derivatives. We could obtain more similar forms by reversing 
the arrow of time and the sign of the gravitational interaction. 
In this sense, therefore, the Lagrangian Bernoulli equation acts 
as a sort of {\em time machine} (cf. Nusser \& Dekel 1992). 
This fact becomes more clear if we think to the fact that, by solving it, 
we are indeed asking how the Lagrangian (i.e. initial) geometry at ${\bf q}$
should modify itself in order to reproduce the Eulerian (i.e. evolved) 
properties of the velocity and density fields at the point ${\bf x}
({\bf q}, \tau)$ as time goes on. 

This equation could be used in principle as an alternative 
Lagrangian formulation of Newtonian theory, whose fundamental 
variables would be the velocity potential $\Phi_v^{(L)}$, the gravitational 
potential $\varphi_g^{(L)}$ and the metric tensor 
$\bar \gamma_{\alpha\beta}$. This approach could be useful, in particular, 
in order to obtain new self--consistent approximation schemes to the 
non--linear evolution of dust in the Lagrangian frame. To 
this aim, however, we need two more equations to close the system. 
These can be provided by the Lagrangian Poisson equation above, Eq.(91), and 
by the very definition of $\bar \vartheta_{\alpha\beta} = 
\bar \gamma_{\alpha\gamma}
\bar \vartheta^\gamma_{~\beta}$, which implies 
\be
\Phi_{v|\alpha\beta}^{(L)} = {1 \over 2} {\bar \gamma_{\alpha\beta}}' \;.
\ee

Of course, the Lagrangian scalars $\Phi_v^{(L)}$ and 
$\varphi_g^{(L)}$ are related to their 
Eulerian counterparts by a simple coordinate transformation, namely 
$\Phi_v^{(L)} ({\bf q}, \tau) = \Phi_v^{(E)} ({\bf x} ({\bf q}, \tau), 
\tau)$ and $\varphi_g^{(L)} ({\bf q}, \tau) = \varphi_g ^{(E)} 
({\bf x} ({\bf q}, \tau), \tau)$. 

%%%%%%%%%%%%%%%%%%%%%%%%%%%%%%%%%%%%%%%%%%%%%%%%%%%%%%%%%%%%%%%%%%%%%%%
\section{Post--Newtonian approximation }
%%%%%%%%%%%%%%%%%%%%%%%%%%%%%%%%%%%%%%%%%%%%%%%%%%%%%%%%%%%%%%%%%%%%%%%

Having examined all the aspects of our formalism in the Newtonian 
limit, we are now ready to proceed to the next perturbative order in $1/c^2$. 
The PN terms $\frac {1}{c^2} w^{(PN)}_{\alpha\beta}$ in Eq.(56) should be 
thought as small perturbations superposed on a Newtonian background
$\bar \gamma_{\alpha\beta}$. 
The fact that the three--metric in the Newtonian limit is that of
Euclidean space in time--dependent curvilinear coordinates $q^\alpha$, implies
that we can apply most of the standard tools of linear perturbation theory in
a flat spatial background. In 
particular, we can classify our PN metric perturbations as 
scalar, vector and tensor modes, as usual. 

We then write
\be
w^{(PN)}_{\alpha\beta} = 
\chi^{(PN)} {\bar \gamma}_{\alpha\beta} + \zeta^{(PN)}_{|\alpha\beta} +
{1 \over 2} \bigl( \xi^{(PN)}_{\alpha | \beta} + \xi^{(PN)}_{\beta | \alpha} 
\bigr) + \pi^{(PN)}_{\alpha\beta} \;, 
\ee
with
\be
{\xi^{(PN)}}^\alpha_{~~|\alpha}= {\pi^{(PN)}}^\alpha_{~\alpha} = 
{\pi^{(PN)}}^\alpha_{~\beta|\alpha}=0 \;,
\ee
where greek indices after a single vertical bar, or nabla operators with a 
greek index, denote covariant differentiation in the 
Newtonian background three--space with metric ${\bar \gamma}_{\alpha\beta}$. 
In the above decomposition $\chi^{(PN)}$ and $\zeta^{(PN)}$ represent PN 
scalar modes, $\xi^{(PN)}_\alpha$ PN vector modes and 
$\pi^{(PN)}_{\alpha\beta}$ PN tensor ones 
(indices being raised by the contravariant background three--metric). 
We deliberately used the same symbols as in Section 2.4, in order to 
emphasize the analogy with the linear problem. 
Some of these PN modes, namely $\chi^{(PN)}$ and 
$\pi^{(PN)}_{\alpha\beta}$, also have a non--vanishing linear counterpart,
as noticed in Section 2.4 (actually the linear part of 
$\pi^{(PN)}_{\alpha\beta}$ appears as a gauge mode in the equations), 
while others, namely $\zeta^{(PN)}$, and $\xi_\alpha^{(PN)}$ are intrinsically 
non--linear. Unlike linear perturbation theory in a FRW background, 
metric perturbations of different rank do not decouple: this is because 
our time--dependent Newtonian background enters the 
equations not only through the metric $\bar \gamma_{\alpha\beta}$, but also 
through the peculiar velocity--gradient tensor 
$\bar \vartheta^\alpha_{~\beta}$, which also contains scalar, vector and tensor 
modes. This fact leads to non--linear scalar--vector, 
scalar--tensor and vector--tensor mode mixing, which also explains why we
had to account for the vector modes $\xi^{(PN)}_\alpha$ in the expansion of 
$w^{(PN)}_{\alpha\beta}$, in spite of the irrotational character of our fluid 
motions. Actually, that vector modes appear in the non--linear 
evolution of an irrotational fluid in Lagrangian coordinates is well known 
also in the Newtonian framework (see Buchert 1994; Catelan 1995). 

As in every perturbative calculation, some of the equations 
have the property to mix different perturbative orders. This is of course 
necessary in order to make the $n$--th order coefficients of the expansion 
calculable in terms of those of order $n-1$. In our case the energy constraint
and the extrinsic curvature evolution equation (which at the Newtonian level 
implies ${\cal R}^\alpha_{~\beta}(\bar \gamma)=0$) play this role. 
Therefore we assume that the Newtonian metric and its derivatives are known 
by solving the Raychaudhuri equation and the momentum constraint, 
and we calculate the PN metric perturbations in terms of them. 

Let us first compute the tensor ${\cal P}^\alpha_{~\beta} \equiv 
4 {\cal R}^\alpha_{~\beta} - \bigl( {\cal R} + 2 \kappa \bigr) 
\delta^\alpha_{~\beta}$ to first order in $1/c^2$. We obtain 
\be
c^2 {\cal P}^\alpha_{~\beta} (w^{(PN)})  = - 2 \biggl( 
\nabla^2 {\pi^{(PN)}}^\alpha_{~\beta} + {\chi^{(PN)}_|}^\alpha_{~\beta} 
\biggr) \;, 
\ee
where now $\nabla^2 ( \cdot ) \equiv {( \cdot )_|}^\alpha_{~\alpha}$.

In the energy constraint only the scalar $\chi$ enters, 
\be
\nabla^2 \chi = {1 \over 2} \biggl( \bar \vartheta^2 - \bar 
\vartheta^\mu_{~\nu} \bar \vartheta^\nu_{~\mu} \biggr) + 
2 {a' \over a} \bar \vartheta - 8 \pi G a^2 \varrho_b 
\bigl( \bar \gamma^{-1/2} - 1 \bigr) \;, 
\ee
where, here and from now on, we have dropped the superscript (PN) 
on PN terms. 
One can also obtain an equation for $\chi$ from the trace of the evolution 
equation, Eq.(22), which is however equivalent to the latter, 
thanks to the Newtonian Raychaudhuri equation. 

The tensor perturbations $\pi^\alpha_{~\beta}$ are instead determined via
the evolution equation, Eq.(22) (actually from its trace--free part), 
\be
\nabla^2 \pi^\alpha_{~\beta} = 2 {\bar \vartheta}^{\alpha~'}_{~\beta}
+ 2 \biggl( 2 {a' \over a} + \bar \vartheta \biggr) 
\bar \vartheta^\alpha_{~\beta} - 
{1 \over 2} \delta^\alpha_{~\beta} 
\biggl( \bar \vartheta^2 - 
\bar \vartheta^\mu_{~\nu} \bar \vartheta^\nu_{~\mu} \biggr) - 
{\chi_|}^\alpha_{~\beta} \;. 
\ee

A by--product of the latter equation is that linear tensor modes, which in 
the $c \to \infty$ limit appear as harmonic functions (i.e. pure gauge modes), 
do not contribute to the r.h.s., i.e. to the Newtonian evolution of the 
system, as expected. 

In order to get an equation for the tensor modes decoupled from the scalar 
mode $\chi$ we can resort to the equations obtained in Section 3.3 above. 
To this aim we define the auxiliary Newtonian potential $\Psi_v^{(L)}$, through 
\be
\nabla^2 \Psi_v^{(L)} = - {1 \over 2} \biggl( \bar \vartheta^2 - \bar 
\vartheta^\mu_{~\nu} \bar \vartheta^\nu_{~\mu} \biggr) \;.
\ee

Using this definition in Eq.(98), we obtain 
\be
\chi = - \Psi_v^{(L)} + 2 {a' \over a} \Phi_v^{(L)}- 2 \varphi_g^{(L)} \; 
\ee
(which, at the linear level reduces to the expression of Section 2.4, 
$\chi_0= - \frac{10}{3c^2} \varphi_0$).  
Using this expression in Eq.(99) and replacing 
${\bar \vartheta}^{\alpha~'}_{~\beta}$ from Eq.(90) we obtain
\be
\nabla^2 \pi^\alpha_{~\beta} = \biggl( \nabla^\alpha \nabla_\beta 
+ \delta^\alpha_{~\beta} \nabla^2 \biggr) \Psi_v^{(L)} + 
2 \biggl( \bar \vartheta \bar \vartheta^\alpha_{~\beta} - 
\bar \vartheta^\alpha_{~\gamma} 
\bar \vartheta^\gamma_{~\beta} \biggr) \;, 
\ee
which has the significant advantage of being explicitly second 
order (in any possible perturbative approach). 
This equation is one of the most important results of this paper: 
it gives (in the so--called {\em near zone}) the amount of gravitational waves 
emitted by non--linear cosmological perturbations, 
evolved within Newtonian gravity. In other terms, this
equation, which is only applicable on scales well inside 
the horizon, describes gravitational waves produced by an 
inhomogeneous Newtonian background. 
At first sight it may appear surprising that gravitational waves already 
appear in a PN calculation, whilst in the standard calculations 
in non--comoving gauges one finds them at the PPN level. 
This is indeed a peculiarity (actually an 
advantage) of the Lagrangian coordinates, where all orders are scaled by two 
powers of $c$: for instance, the Newtonian terms already appear at the zeroth 
order, whereas in the longitudinal gauge the gravitational 
potential carries a $1/c^2$ factor. 

This formula can be compared with the PN limit of Eq.(3) in Section 1, 
to which it actually reduces (up to an already mentioned numerical factor) 
if the $\vartheta^\alpha_{~\beta}$ are 
calculated from linear theory and in an Einstein--de Sitter model. We have 
$\bar \vartheta^\alpha_{~\beta} ({\bf q}, \tau)= D'(\tau) 
{\Phi_{0,}}^\alpha_{~\beta}({\bf q})$, 
with $D(\tau)$ the growing mode solution of Eq.(52) ($D(\tau) \propto a(\tau) 
\propto \tau^2$ in the Einstein--de Sitter case) and $\Phi_0$ defined in 
Section 3.2. Therefore 
\be
\nabla_q^2 \pi^\alpha_{~\beta} = {D'}^2 \biggl[ 
{\Psi_{0,}}^\alpha_{~\beta} + \delta^\alpha_{\beta} 
\nabla_q^2 \Psi_0 + 2 \biggl( {\Phi_{0,}}^\alpha_{~\beta} 
\nabla_q^2 \Phi_0 - {\Phi_{0,}}^\alpha_{~\gamma}  
{\Phi_{0,}}^\gamma_{~\beta} \biggr) \biggr] \;, 
\ee
where the symbol $\nabla^2_q$ indicates the standard (Euclidean) form of the 
Laplacian in Lagrangian coordinates and $\Psi_0 \equiv \Psi_v(\tau_0)$ and 
indices are raised by the Kronecker symbol. 

To completely determine the PN metric perturbations we still need 
the scalar mode $\zeta$ and the vector modes $\xi_\alpha$, which can be
computed through the momentum constraint and the Raychaudhuri equation.  
We then need $\vartheta^\alpha_{~\beta}$ to PN order; it reads
\be
\vartheta^\alpha_{~\beta} = \bar \vartheta^\alpha_{~\beta} + {1 \over c^2}
\biggl( \bar \vartheta^\alpha_{~\gamma} w^\gamma_{~\beta} - 
\bar \vartheta^\gamma_{~\beta} w^\alpha_{~\gamma} + 
{1 \over 2} {w^\alpha_{~\beta}}' \biggr) \;. 
\ee
By replacing this expression into the momentum constraint, Eq.(21), we obtain
\begin{eqnarray}
\nonumber
\bigl({w^\alpha_{~\beta}}' \bigr)_{|\alpha} - w'_{,\beta}
+ 2 \bar \vartheta_{,\alpha} w^\alpha_{~\beta}
- 2 \bar \vartheta^\gamma_{~\beta|\alpha} w^\alpha_{~\gamma}
+ 2 \bar \vartheta^\gamma_{~\alpha} w^\alpha_{~\beta|\gamma}
- 2 \bar \vartheta^\gamma_{~\beta} w^\alpha_{~\gamma|\alpha}
- \bar \vartheta^\gamma_{~\alpha} w^\alpha_{~\gamma|\beta}
+ \bar \vartheta^\alpha_{~\beta} w_{,\alpha} = 
\\
= - 4 \kappa_N
\Phi_{v,\beta}^{(L)} \;,
\end{eqnarray}
where the term on the r.h.s. comes from the relation
$\bar \vartheta^\alpha_{~\beta|\alpha} - \bar \vartheta_{,\beta} =
2 \kappa  \Phi_{v,\beta}^{(L)}$, which can be easily derived by 
expanding the Jacobi identity
${\Phi_{v||}^{(L)}}^\alpha_{~~\beta\alpha}-
{\Phi_{v||}^{(L)}}^\alpha_{~~\alpha\beta} = \Phi_{v,\alpha}^{(L)}
{\cal R}^\alpha_{~\beta}$, in powers of $1/c^2$.

In order to write the last equation in terms of the various PN perturbation 
modes, the Newtonian identity ${\bar \Gamma}^{\mu~'}_{\nu\rho} = 
\bar \vartheta^\mu_{~\nu|\rho}$ is also useful, implying
\be
\bigl({w^\alpha_{~\beta}}' \bigr)_{|\alpha} = 
\bigl(w^\alpha_{~\beta|\alpha} \bigr)' - \bar \vartheta_{,\alpha}
w^\alpha_{~\beta} + \bar \vartheta^\gamma_{~\alpha|\beta} w^\alpha_{~\gamma} 
\;.
\ee

By replacing the expansion of $w^\alpha_{~\beta}$ into this equation we 
finally get 
\begin{eqnarray}
\nonumber
2 {\chi_{,\beta}}' + 
\bar \vartheta \chi_{,\beta} - 
3 \bar \vartheta^\alpha_{~\beta} \chi_{,\alpha} - 
\bar \vartheta_{,\alpha} \pi^\alpha_{~\beta} + 
\bar \vartheta^\gamma_{~\beta|\alpha} \pi^\alpha_{~\gamma} - 
2 \bar \vartheta^\alpha_{~\gamma} \pi^\gamma_{~\beta|\alpha} 
+ \bar \vartheta^\alpha_{~\gamma} \pi^\gamma_{~\alpha|\beta} -
\\
\nonumber
 - \bar \vartheta_{,\alpha} {\zeta_|}^\alpha_{~\beta} + 
\bar \vartheta^\gamma_{~\beta|\alpha} {\zeta_|}^\alpha_{~\gamma} -
\bar \vartheta^\alpha_{~\gamma} {\zeta_|}^\gamma_{~\beta\alpha} + 
\bar \vartheta^\alpha_{~\beta} {\zeta_|}^\gamma_{~\alpha\gamma} - 
{1 \over 2} \bigl(\nabla^2 \xi_\beta \bigr)' - 
{1 \over 2} \bar \vartheta_{,\alpha} \xi^\alpha_{~|\beta} -
\\
 - {1 \over 2} \bar \vartheta_{,\alpha} \xi_{\beta|}^{~~\alpha} + 
\bar \vartheta^\gamma_{~\beta|\alpha} \xi^\alpha_{~|\gamma} - 
\bar \vartheta^\alpha_{~\gamma} {\xi_{\beta|}}^\gamma_{~\alpha} +
\bar \vartheta^\alpha_{~\beta} \nabla^2 \xi_\alpha = 4 \kappa_N
{\Phi_{v|}^{(L)}}_\beta \;,
\end{eqnarray}
which, at the linear level, reduces to Eq.(47) above.

One can verify that the three--divergence of the last equation reduces to an 
indentity, therefore, in order to completely determine the three 
remaining PN modes $\zeta$ and $\xi^\alpha$ we need one more equation. This
is in fact provided by the PN Raychaudhuri equation, which reads 
\be
w''+ {a' \over a} w' + 2 \bar \vartheta^\mu_{~\nu} {w^\nu_{~\mu}}' = 
4\pi G a^2 \varrho_b \bigl( 1 + \bar \delta \bigr) \bigl( w - w_0 \bigr) \;,
\ee
having assumed $\delta_0^{(PN)}=0$, as suggested by linear theory. 
By replacing the expansion of $w^\alpha_{~\beta}$ into this equation we get
\begin{eqnarray}
\nonumber
\bigl(3\chi + \nabla^2 \zeta\bigr)'' + {a' \over a} \bigl(3\chi + \nabla^2 
\zeta\bigr)' + 2 \bar \vartheta \chi' + 2 \bar \vartheta^\mu_{~\nu} 
\bigl( {\zeta_|}^\nu_{~\mu} + \xi^\nu_{~|\mu} \pi^\nu_{~\mu} \bigr)' 
= 
\\
= 4\pi G a^2 \varrho_b \bigl( 1 + \bar \delta \bigr) \biggl[3\bigl(\chi-\chi_0)
+ \nabla^2 \zeta \biggr] \;, 
\end{eqnarray}
where we have also set $w_0=3\chi_0$, in agreement with linear theory. 

Unfortunately, we have not been able to further simplify these
equations, which nevertheless show that $\zeta$ and $\xi^\alpha$ are 
implicitly determined by the Newtonian quantities, once $\chi$ and 
$\pi^\alpha_{~\beta}$ have been computed. There is a caveat concerning 
second--order perturbation theory, where $\zeta$ is left undetermined by the 
above equations. Nevertheless, such an 
ambiguity is completely removed by going to the PPN energy constraint, which 
can be easily shown to fix $w$ and then $\zeta$ without explicit knowledge of 
the truly PPN coefficients. Let us complete this discussion by reporting
the PPN energy constraint. 
Defining a PPN metric perturbation $w^{(PPN)}_{\alpha\beta}/c^4$, one gets 
\be
\bigl( \bar \vartheta + 2 {a' \over a} \bigr) w' - {\bar \vartheta}^\mu_{~\nu}
w^{\nu~'}_{~\mu} + {\cal R}^{(PPN)} = - 8\pi G a^2 \varrho_b
\bigl( 1 + {\bar \delta} \bigr) \bigl( w - w_0 \bigr) \;,
\ee
with ${\cal R}^{(PPN)}$ the PPN conformal Ricci scalar, 
\begin{eqnarray}
\nonumber
{\cal R}^{(PPN)} = - 2 \kappa_N w + {w^{(PPN)}}^{\mu\nu}_{~~~|\mu\nu} -
\nabla^2 w^{(PPN)} + w^\mu_{~\nu} \bigl( {w_|}^\nu_{~\mu} +
\nabla^2 w^\nu_{~\mu}
- 2 w^{\nu\gamma}_{~~~|\mu\gamma} \bigr) + 
\\
+ {3 \over 4} {w^\mu_{~\nu|}}^\gamma w^\nu_{~\mu|\gamma} - 
{1 \over 2} {w^\mu_{~\nu|}}^\gamma {w_{\mu\gamma|}}^\nu -
{1 \over 4} \bigl( w_{,\nu} - 2 w^\mu_{~\nu|\mu} \bigr)
\bigl( w_|^{~\nu} - 2 w^{\gamma\nu}_{~~~|\gamma} \bigr) \;.
\end{eqnarray}

%%%%%%%%%%%%%%%%%%%%%%%%%%%%%%%%%%%%%%%%%%%%%%%%%%%%%%%%%%%%%%%%%%%%%%%
\subsection{Fluid--flow approach in the Newtonian limit }
%%%%%%%%%%%%%%%%%%%%%%%%%%%%%%%%%%%%%%%%%%%%%%%%%%%%%%%%%%%%%%%%%%%%%%%

We are now ready to discuss the fluid--flow approach presented in Section 2.2, 
within the Newtonian approximation.  The reason why this discussion has been 
included in this Section is that, as we shall see, some of the relevant tensors 
must be computed at the PN order in order to provide the correct Newtonian 
evolution of the system. 

We just have to discuss the order in our $1/c^2$ expansion at which the 
various tensors enter the equations of Section 2.2. 
It is immediately clear that the mass continuity equation, Eq.(25), the 
Raychaudhuri equation, Eq.(26), and the shear evolution equation, Eq.(27), 
where no explicit powers of $c$ appear, just keep their form, once the 
various tensors are replaced by their Newtonian counterparts. So, we have 
\be
\bar \delta' + \bigl(1 + {\bar \delta} \bigr) {\bar \vartheta} = 0 \;,
\ee
\be
{\bar \vartheta}' + {a' \over a} {\bar \vartheta} + {1 \over 3} 
{\bar \vartheta}^2 + {\bar \sigma}^\alpha_{~\beta} 
{\bar \sigma}^\beta_{~\alpha} + 4 \pi G a^2 \varrho_b {\bar \delta} = 0 \;,
\ee
and 
\be 
{\bar \sigma}^{\alpha~'}_{~\beta} + {a' \over a} {\bar \sigma}^\alpha_{~\beta}
+ {2 \over 3} {\bar \vartheta} {\bar \sigma}^\alpha_{~\beta} + 
{\bar \sigma}^\alpha_{~\gamma} {\bar \sigma}^\gamma_{~\beta} - {1 \over 3} 
\delta^\alpha_{~\beta}
{\bar \sigma}^\gamma_{~\delta} {\bar \sigma}^\delta_{~\gamma} + 
{\bar {\cal E}}^\alpha_{~\beta} = 0 \;. 
\ee

On the other hand, by its very definition, Eq.(28), the electric tide contains 
a contribution coming from the PN terms $\chi$ and $\pi^\alpha_{~\beta}$ 
because of the spatial curvature terms, 
\be
{\bar {\cal E}}^\alpha_{~\beta} = {1 \over 3} \delta^\alpha_{~\beta} 
{\bar \sigma}^\mu_{~\nu} {\bar \sigma}^\nu_{~\mu} + {1 \over 3} 
{\bar \vartheta } {\bar \sigma}^\alpha_{~\beta} + {a' \over a} 
{\bar \sigma}^\alpha_{~\beta} - 
{\bar \sigma}^\alpha_{~\gamma} {\bar \sigma}^\gamma_{~\beta} 
- {1 \over 2} \biggl[ \nabla^2 \pi^\alpha_{~\beta} + \biggl(\nabla^\alpha
\nabla_\beta - {1 \over 3} \delta^\alpha_{~\beta} \nabla^2 \biggr) \chi 
\biggr] \;.
\ee
It is however immediate to realize that, once the expressions of this section 
for the PN tensors $\chi$ and $\pi^\alpha_{~\beta}$ are used, one recovers 
the simpler form, 
\be
{\bar {\cal E}}^\alpha_{~\beta} = \biggl( \nabla^\alpha \nabla_\beta -
{1 \over 3} \delta^\alpha_{~\beta} \nabla^2 \biggr) \varphi_g^{(L)} \;,
\ee
which, in Eulerian coordinates reduces to the standard 
form 
\be
{\bar {\cal E}}_{AB} = \varphi_{g,AB} - \frac{1}{3} \delta_{AB} \nabla_x^2 
\varphi_g \;. 
\ee

On the other hand, if we replace in Eq.(33), the Newtonian peculiar 
velocity--gradient tensor, 
we obtain the well--known result (e.g. Ellis 1971) that the magnetic tensor 
identically vanishes in the Newtonian limit. This can be very easily 
shown by either applying the formalism of Section 3.3, i.e. writing 
$\bar \vartheta^\alpha_{~\beta}$ through covariant derivatives of the 
Lagrangian velocity potential, or by writing the same tensor 
in terms of the Jacobian matrix of Section 3.1. The physics underlying this 
result is the conformal flatness of the Newtonian spatial sections, implying 
the commutation of spatial covariant derivatives. A simple consequence of this 
fact is that, 
at the Newtonian level, the ${\rm div} ~{\cal E}$ constraint, Eq.(34), reduces
to 
\be
{\bar {\cal E}}^\alpha_{~\beta|\alpha} = {8 \pi G \over 3} a^2 \varrho_b 
{\bar \delta}_{,\beta} \;,
\ee
which, owing to our expression for ${\bar {\cal E}}$, turns out to be just the 
gradient of the Lagrangian Poisson equation, Eq.(91), namely 
\be
\nabla^2 \varphi_g^{(L)} = 4 \pi G a^2 \varrho_b {\bar \delta} \;.
\ee

Let us now come to the tide evolution equation, Eq.(30). 
In that evolution equation the circulation of the magnetic tensor is 
multiplied by 
$c^2$, which means that the PN part of ${\rm curl} ~{\cal H}$ is the source of
non--locality in the Newtonian electric tide evolution equation. 
On the other hand, if we look at the magnetic tide evolution equation, which 
starts to be non--trivial at the PN order, we see that 
${\rm curl} ~{\cal E}$ is consistently a PN quantity. 

The Newtonian meaning of the momentum constraint, Eq.(32), has been already 
discussed in Section 3. 
Also interesting is the ${\rm div} ~{\cal H}$ constraint, Eq.(35), 
telling us that the general non--vanishing of the PN 
magnetic tensor (see also Lesame, Ellis \& Dunsby 1996), implies that the 
Newtonian shear and electric tide do not commute, i.e. they have different 
eigenvectors (viceversa, their non--alinement causes a non--zero 
${\rm div} ~{\cal H}$). Another possible 
version of this result is that the ratio of the velocity potential to 
the gravitational potential beyond the linear regime becomes 
space--dependent. 

To summarize our results, we can say that within the Newtonian approximation 
the fluid--flow approach in Lagrangian coordinates can be formulated in terms 
of mass continuity, Raychaudhuri and shear evolution equations plus 
the Newtonian ${\rm div} ~{\cal E}$ constraint, which closes the system, 
provided we remind the circulation--free character of the electric tide in this 
limit. Of course the direct use of a constraint to close the system of 
evolution equations, has the disadvantage of breaking the intrinsic 
hyperbolicity of the GR set of evolution equations, so that the entire 
method looses its basic feature. No ways out: this is the price to pay to the 
intrinsic non--causality of the Newtonian theory [see also Ellis (1990)]. 

The above discussion on the role of the PN magnetic tidal tensor, 
as causing non--locality in the Newtonian fluid--flow evolution equations, 
completely agrees with a similar analysis by Kofman \& Pogosyan (1995). 
The only variant is that we obtained our results directly in Lagrangian 
space, while they worked in non--comoving (i.e. Eulerian) coordinates. 
A different point of view on the subject is expressed by 
Bertschinger \& Hamilton (1994), according to which the magnetic part of the 
Weyl tensor is non--vanishing already at the Newtonian level. According to 
Kofman \& Pogosyan (1995) the difference might be ``semantic"; 
most important, there is general agreement on the 
fundamental fact that the Newtonian tide evolution is affected by 
non--local terms. 

%%%%%%%%%%%%%%%%%%%%%%%%%%%%%%%%%%%%%%%%%%%%%%%%%%%%%%%%%%%%%%%%%%%%%%%
\subsection{Post--Newtonian tensor modes within a collapsing homogeneous 
ellipsoid }
%%%%%%%%%%%%%%%%%%%%%%%%%%%%%%%%%%%%%%%%%%%%%%%%%%%%%%%%%%%%%%%%%%%%%%%

The PN expression for $\pi_{\alpha\beta}$, Eq.(102), has the relevant 
feature of being non--local, through the presence of the scalar 
$\Psi_v$. A simpler way to deal with this problem is to transform the equation
in Eulerian form, where it is easier to deal with the Laplacian operator
$\nabla_x^2$ (which has there the standard Euclidean form), 
obtain the Eulerian gravitational--wave tensor $\pi_{AB}$ and then go back 
to the Lagrangian expression through 
$\pi_{\alpha\beta} ({\bf q}, \tau) = {\bar {\cal J}}^A_{~~\alpha} 
{\bar {\cal J}}^B_{~~\beta} \pi_{AB}({\bf x}({\bf q},\tau), \tau)$. 
We immediately obtain the Eulerian expression
\be
\nabla^2_x \pi_{AB} = \Psi^{(E)}_{v,AB} + \delta_{AB} \nabla_x^2 \Psi_v^{(E)}
+ 2 \biggl( \bar \vartheta \bar \vartheta_{AB} - 
\bar \vartheta_{AC} 
\bar \vartheta^C_{~~B} \biggr) \;, 
\ee
with 
\be
\nabla_x^2 \Psi_v^{(E)} = - {1 \over 2} 
\biggl( \bar \vartheta^2 - 
\bar \vartheta^A_{~B}  
\bar \vartheta^B_{~A} \biggr) \;, 
\ee
which generally allows a simpler derivation of $\pi_{AB}$, given the 
(gradients of the) velocity potential. For a general homogeneous and isotropic 
random field $\Phi_v$, for instance, $\pi_{AB}$ can be obtained by a simple 
convolution in Fourier space. 
Nevertheless, we would like to obtain here an analytic estimate of this 
tensor, in some simple cases. What we need is a model for non--spherical and
non--planar collapse. The simplest model we can 
figure out is that of a homogeneous ellipsoid with uniform internal overdensity 
$\delta(\tau)$ with respect to a FRW background, of density $\varrho_b(\tau)$
and scale factor $a(\tau)$, in which it is embedded 
(White \& Silk 1979; Peebles 1980). Calling
$R_A(\tau) \equiv a(\tau)X_A(\tau)$, $A=1,2,3$, the physical length of the 
three axes, the peculiar gravitational potential within the ellipsoid is given 
by the simple expression 
\be
\varphi_g({\bf x}, \tau) = \pi G a^2 \varrho_b \delta \sum_A\alpha_A 
x_A^2 \;,
\ee
The dimensionless structure constants $\alpha_A$ are defined in terms 
of the three $X_A$ through 
\be
\alpha_A = X_1 X_2 X_3 \int_0^\infty ds (X_A^2 + s )^{-1} \prod_B
(X_B^2 + s)^{-1/2} 
\ee
and are normalized so that $\sum_A\alpha_A=2$. 
In the particular case of oblate or prolate spheroids one can get explicit 
expressions for the $\alpha_A$ in terms of an eccentricity parameter (e.g.
Kellogg 1953). The simplest non--trivial case, however, is that of an infinite 
cylinder, for which the two non--vanishing structure constants have the value 
$1$ at any time. 
The intrinsic self--similarity of the equations of motion for fluid elements 
within the object implies that the overall shape and homogeneity are preserved 
at all times. One then usually makes the reasonable approximation that also 
the universe outside the ellipsoid stays uniform. In such a case, the 
ellipsoid axes obey the equation 
\be
X_A'' + {a' \over a} X_A' = -2 \pi G a^2 \varrho_b\delta ~\alpha_A X_A \;, 
\ee 
while mass conservation implies 
\be
(1 + \delta) X_1 X_2 X_3 = const \;.
\ee
The peculiar velocity--gradient tensor has eigenvalues $\vartheta_A= X'_A/X_A$ 
(which remain unchanged in Lagrangian coordinates). These determine 
the Eulerian potential $\Psi_v^{(E)}$, through the equation
\be
\nabla_x^2 \Psi_v^{(E)} = - \sum_A \vartheta_A \vartheta_{A-1} \;,
\ee
where we adopt a notation such that $A-1=3$ if $A=1$ and $A+1=1$ if $A=3$. 
Accounting for the ellipsoidal symmetry of the problem, this equation is 
solved by
\be
\Psi_v^{(E)} = - {1 \over 4} \sum_A \vartheta_A \vartheta_{A-1} 
\sum_B \alpha_B x_B^2 \;.
\ee
Replacing this solution in Eq.(120) gives 
\be
\nabla_x^2 \pi_A = \mu_A(\tau) \equiv - {1 \over 2} \bigl(
\vartheta_A \vartheta_{A-1} + \vartheta_A \vartheta_{A+1} 
+ \vartheta_{A-1} \vartheta_{A+1} \bigr) \alpha_A + 
\bigl( \vartheta_A \vartheta_{A-1} + \vartheta_A \vartheta_{A+1} 
- \vartheta_{A-1} \vartheta_{A+1} \bigr) \;,
\ee
where $\pi_A \equiv \pi_{AA}$ (no summation over repeated indices is 
understood) indicates a diagonal component, and $\sum_A\mu_A=0$. This equation 
is solved by 
\be
\pi_A = {1 \over 4} \mu_A \sum_B \alpha_B x_B^2 \;.
\ee
The off--diagonal components are instead harmonic functions, 
\be
\nabla_x^2 \pi_{AB}= 0 \;, \ \ \ \ \ \ \ A \neq B \;. 
\ee
These equations must be solved accounting for the transversality 
condition $\sum_A \pi_{AB,A}=0$, which gives 
\be
\pi_{AB} = {1 \over 4} \nu_{AB} x_A x_B \;, \ \ \ \ \ \ \ A \neq B 
\ee
(no summation over repeated indices), with $\nu_{AB}=\nu_{BA}$ and
\be
\begin{array}{c}
\nu_{12}= \mu_3 \alpha_3-\mu_1 \alpha_1-\mu_2 \alpha_2  \\[0,3cm]
\nu_{13}= \mu_2 \alpha_2-\mu_1 \alpha_1-\mu_3 \alpha_3  \\[0,3cm]
\nu_{23}= \mu_1 \alpha_1-\mu_2 \alpha_2-\mu_3 \alpha_3 \;. \\[0,3cm]
\end{array}
\ee
These formulae are completely general and do not contain approximations, 
(apart from those implicit in the homogeneous ellipsoid model). 
The results of a numerical integration of Eqs.(123), (124) and (125) are shown 
in Figure 1, for the collapse of a triaxial homogeneous ellipsoid. 

One might also use them to get the gravitational--wave emission outside the 
object (i.e. in the {\em wave zone}), by suitable matching with the interior 
solution. Far away from the body, one could, of course, also obtain the emitted
gravitational radiation in terms of the quadrupole moments 
of our homogeneous ellipsoid (e.g. Landau \& Lifshitz 1980, Sects. 41 and 105)
and recover the usual $1/c^5$ dependence of the radiated energy. In the 
wave zone, moreover, the transverse and traceless 
character of the $\pi_{AB}$ would allow to define the two 
polarization states of the graviton. These problems will not be considered 
here; it is nevertheless important to realize that our formalism is completely 
consistent with the qualitative expectations of the quadrupole approximation, 
as it implies identically vanishing tensor modes in the case of 
spherical ($\alpha_1=\alpha_2=\alpha_3=2/3$) and plane--parallel 
($\alpha_1=\alpha_2=0$, $\alpha_3=2$) collapse. 

What we are interested in here is the behaviour of the PN tensor modes when 
the object is close to collapse. 
Of course, the set formed by Eq.(123), Eq.(124) and Eq.(125) could 
be integrated numerically to get the time evolution for the axes $X_A$, 
the eigenvalues $\vartheta_A$, and the structure constants 
$\alpha_A$ (e.g. White \& Silk 1979). 
To catch the qualitative behaviour close to collapse, however, we can safely 
apply the Zel'dovich approximation, which, for the evolution of the axes, 
yields 
\be
X_A(\tau) = X_A(\tau_0) (1 + D(\tau) \lambda_A ) \;,
\ee
with $\lambda_A = - \frac{\delta_0}{2 D_0} \alpha_{A}(\tau_0)$. 
These expressions should then be replaced into the definition of the $\alpha_A$
to get them self--consistently. Nevertheless, according to White \& Silk 
(1979) a rough estimate is obtained by simply neglecting the time--dependence 
of the $\alpha_A$. 
One can immediately derive the Jacobian eigenvalues ${\bar {\cal J}}_\alpha 
= 1 + D \lambda_\alpha$, with $A=\alpha$, and those of the peculiar 
velocity--gradient tensor $\bar \vartheta_\alpha = D'\lambda_\alpha/
(1 + D \lambda_\alpha)$.

These expressions can then be replaced into the previous equations to get 
the Lagrangian relations 
\be
\pi_\alpha \equiv \pi_{\alpha\alpha} ({\bf q}, \tau) = {1 \over 4} \mu_\alpha
{\bar {\cal J}}_\alpha^2 \sum_\beta \alpha_\beta {\bar {\cal J}}_\beta^2 
q_\beta^2 \;,
\ee
for the diagonal components, and 
\be
\pi_{\alpha\beta} ({\bf q}, \tau) = {1 \over 4} \nu_{\alpha\beta} 
{\bar {\cal J}}_\alpha^2 {\bar {\cal J}}_\beta^2 q_\alpha q_\beta \;,
\ \ \ \ \ \ \alpha \neq \beta \;,
\ee
for the off--diagonal ones, with 
\be
\begin{array}{c}
\ds{ \mu_\alpha =\frac{{D'}^2}{\bar {\cal J}} \biggl[{1 \over 2} 
        \biggl(\lambda_{\alpha-1}
        \lambda_{\alpha+1}+\lambda_\alpha\lambda_{\alpha-1}+
        \lambda_\alpha\lambda_{\alpha+1}+3D\lambda_\alpha
        \lambda_{\alpha-1}\lambda_{\alpha+1} \biggr) \alpha_\alpha+ } \\[0,3cm]
\ds{ +  \biggl(\lambda_\alpha
        \lambda_{\alpha-1}+\lambda_\alpha\lambda_{\alpha+1}-
        \lambda_{\alpha-1}\lambda_{\alpha+1}+D\lambda_\alpha
        \lambda_{\alpha-1}\lambda_{\alpha+1}\biggr) \biggr]}
\end{array}
\ee
and $\nu_{\alpha\beta}$ calculated from these $\mu_\alpha$ according to 
Eq.(132). 
For the most typical case of pancake collapse, where one Jacobian eigenvalue 
goes to zero first, these expressions also go to zero, like 
${\bar {\cal J}}$. 

At this point we are able to compare the behaviour of these PN tensor modes 
to that of the Newtonian part of the metric, which is 
diagonal with eigenvalues $\bar \gamma_\alpha = {\bar {\cal J}}_\alpha^2$. 
It is then clear that these PN modes vanish more slowly than 
the Newtonian part; their ratio diverges like ${\bar {\cal J}}^{-1}$, i.e.
like the mass density at collapse. 
Using a more refined approximation for the 
axes evolution, such as the one proposed by White \& Silk (1979) or that 
recently proposed by Hui \& Bertschinger (1995), would
not change this qualitative result. Indeed, the numerical integration 
we have performed shows that such a divergence can be even stronger, 
as clearly displayed by the last panel in Figure 1. 

The homogeneous ellipsoid model we have worked out does not allow, 
unfortunately, to distinguish the global collapse from a
shell--crossing singularity, but we may argue that this qualitative 
behaviour would generally apply even at shell--crossing. 

%%%%%%%%%%%%%%%%%%%%%%%%%%%%%%%%%%%%%%%%%%%%%%%%%%%%%%%%%%%%%%%%%%%%%%%
\section{Conclusions}
%%%%%%%%%%%%%%%%%%%%%%%%%%%%%%%%%%%%%%%%%%%%%%%%%%%%%%%%%%%%%%%%%%%%%%%

In this paper we have considered a Lagrangian approach to the evolution
of an irrotational and collisionless fluid in general relativity. The use of 
a synchronous and comoving gauge allowed  to reduce the fundamental variables 
of the system to the six metric tensor components of the spatial hypersurface 
orthogonal to the flow lines. 
Our method was based on a standard $1/c$ expansion of the Einstein and 
continuity equations which led to a new, purely Lagrangian, derivation of 
the Newtonian approximation. One of the most important results in this 
respect is that we obtained a simple and transparent expression for the 
Lagrangian metric; exploiting the vanishing of the spatial 
curvature in the Newtonian limit we were able to write it in terms of the 
displacement vector ${\bf S}({\bf q}, \tau) = {\bf x}({\bf q},\tau)  - 
{\bf q}$, from the Lagrangian coordinate ${\bf q}$ to the Eulerian 
one ${\bf x}$ of 
each fluid element, namely
\be
d s^2 = a^2(\tau) \biggl[ - c^2 d \tau^2 + \delta_{AB} 
\biggl(\delta^A_{~\alpha} + {\partial S^A({\bf q}, \tau) 
 \over \partial q^\alpha} \biggr) 
\biggl(\delta^B_{~\beta} + {\partial S^B({\bf q}, \tau) 
\over \partial q^\beta} \biggr) \biggr] \;.
\ee
The spatial metric is that of Euclidean space in time--dependent 
curvilinear coordinates, consistently with the intuitive notion 
of Lagrangian picture in the Newtonian limit. 
Read this way, the complicated equations of Newtonian gravity in the 
Lagrangian picture become much easier: one just has to deal with the spatial 
metric tensor and its derivatives. The involved matrices appearing in 
the standard formulation are nothing else than the covariant and contravariant 
metric tensor and the spatial Christoffel symbols, appearing in covariant 
derivatives. Moreover, the fact that the spatial Ricci curvature 
vanishes in this limit has the great practical advantage that spatial 
covariant derivatives commute. 

Next, we considered the post--Newtonian corrections to the metric and 
wrote equations for them. In particular, we were able to derive a 
simple and general equation for gravitational--wave emission from 
non--linear structures described through Newtonian gravity. The result is 
expressed in Lagrangian coordinates by Eq.(102), but it can also be given 
the Eulerian form of Eq.(120). These formulae allow to calculate the 
amplitude of the gravitational--wave modes in terms of the velocity 
potential $\Phi_v$, which in turn can be deduced from observational data on 
radial peculiar velocities of galaxies, applying the POTENT technique
(Bertschinger \& Dekel 1989). 

In the standard case, where the cosmological perturbations form 
a homogeneous and isotropic random field, we can obtain a heuristic 
perturbative estimate of their amplitude in terms of the 
{\em rms} density contrast and of the ratio of the typical perturbation scale 
$\lambda$ to the Hubble radius $r_H=c H^{-1}$. One simply has
\be
{\pi_{rms} \over c^2} \sim 
\delta_{rms}^2 \biggl( {\lambda \over r_H} \biggr)^2 \;,
\ee
as it can be easily deduced from Eq.(103), specialized to an Einstein--de 
Sitter model. This effect gives rise to a stochastic background of 
gravitational waves which gets a non--negligible amplitude in 
the so--called {\em extremely--low--frequency} band 
(e.g. Thorne 1995), around $10^{-14}$ --  $10^{-15}$ Hz. 
We can roughly estimate that the present--day closure density of this
gravitational--wave background is 
\be
\Omega_{gw}(\lambda) \sim \delta_{rms}^4 
\biggl( {\lambda \over r_H} \biggr)^2 \;.
\ee 
Note that this background is mostly produced {\em here and now}, so its
energy is not affected by the usual $a^{-4}$ dilution of gravitational 
radiation within the Hubble radius. 
In standard scenarios for the formation of structure in the universe, 
the typical density contrast on scales 
$1$ -- $10$ Mpc implies that $\Omega_{gw}$ is about $10^{-5}$ -- 
$10^{-6}$. We might speculate that such a background would give rise to 
secondary CMB anisotropies on intermediate angular scales: a sort of 
{\em tensor Rees--Sciama effect}. This issue will be considered in 
more detail elsewhere. 

On much smaller scales, where the effect might be even more relevant, 
pressure gradients and viscosity cannot be disregarded anymore and the 
entire formalism needs to be largely modified. 

However, our PN formula also applies to isolated structures, where 
the density contrast can be much higher than the {\em rms} value, 
and, what is most important here, shear anisotropies play a fundamental role,
as it happens in the formation of pancakes. A calculation of 
$\pi_{\alpha\beta}$ in the simple case of a homogeneous ellipsoid 
showed that the PN tensor modes become 
dominant, compared to the Newtonian contributions to the metric tensor, 
during the late stages of collapse, and possibly even in the case of a 
shell--crossing singularity. 
There are a number of important limitations of this result, 
the most important of which is the role that pressure would certainly 
play during the highly non--linear stages. A possible consequence 
could be that pressure gradients halt the growth of anisotropy {\em 
before} our relativistic effects come into play. It is nevertheless 
important to stress that our effect generally contradicts the standard 
paradigm, according to which the smallest scale for the applicability of the 
Newtonian approximation is set by the Schwarzschild radius of the object. 
Such a critical scale is indeed only relevant for nearly spherical collapse, 
whereas our effect becomes important precisely if the collapsing structure 
strongly deviates from sphericity. 
On the other hand, if we consider the dynamics of a collisionless fluid 
as a formal problem on itself, the fact that PN terms dominate over Newtonian 
ones implies that in such a regime the perturbative $1/c$ expansion 
breaks down and one should resort to a fully relativistic approach. 

\section*{Acknowledgments}
The Italian MURST is acknowledged for financial support. We would like to
thank M. Bruni, T. Buchert, L. Hui, R. Kates, S. Mollerach and J. Miller, for 
many useful discussions. 
\clearpage
%%%%%%%%%%%%%%%%%%%%%%%%%%%%%%%%%%%%%%%%%%%%%%%%%%%%%%%%%%%%%
\section*{References}
%%%%%%%%%%%%%%%%%%%%%%%%%%%%%%%%%%%%%%%%%%%%%%%%%%%%%%%%%%%%%

%\bibitem{bi:~~}
\quot{~~~~~~~~~~~}
\smallskip

\quot{Arnowitt R., Deser S., Misner C.W., 1962, in Witten L., ed., 
Gravitation: an Introduction to Current Research. Wiley, New York}
\smallskip

\quot{Bardeen J., 1980, Phys. Rev., D22, 1882}
\smallskip

\quot{Barnes A., Rowlingson R.R., 1989, Classical Quantum Gravity, 6, 949}
\smallskip

\quot{Bertschinger E., Dekel A., 1989, ApJ, 336, L5}
\smallskip

\quot{Bertschinger E., Jain B., 1994, ApJ, 431, 486}
\smallskip

\quot{Bertschinger E., Hamilton A., 1994, ApJ, 435, 1}
\smallskip

\quot{Bouchet F.R., Juszkiewicz R., Colombi S., Pellat R., 1992, ApJ, 394, L5}
\smallskip

\quot{Bruni M., Matarrese S., Pantano O., 1995a, ApJ, 445, 958} 
\smallskip

\quot{Bruni M., Matarrese S., Pantano O., 1995b, Phys. Rev. Lett., 74, 11} 
\smallskip

\quot{Buchert T., 1989, A\&A, 223, 9}
\smallskip

\quot{Buchert T., 1992, MNRAS, 254, 729}
\smallskip

\quot{Buchert T., 1994, MNRAS, 267, 811}
\smallskip

\quot{Buchert T., 1995, to appear in Bonometto S., Primack J. \& Provenzale
A., eds., Proc. Enrico Fermi School, Course CXXXII, Dark Matter in the 
Universe, Varenna 1995. Preprint astro-ph/9509005}
\smallskip

\quot{Buchert T., G\"otz G., 1987, J. Math. Phys., 28, 2714}
\smallskip

\quot{Catelan P., 1995, MNRAS, 276, 115} 
\smallskip

\quot{Catelan P., Lucchin F., Matarrese S., Moscardini L., 1995, MNRAS, 276, 
39} 
\smallskip

\quot{Coles P., Lucchin F., 1995, Cosmology: The Origin and Evolution of 
Cosmic Structure. Wiley, Chichester}
\smallskip

\quot{Coles P., Melott A.L., Shandarin S., 1993, MNRAS, 260, 765}
\smallskip

\quot{Croudace K.M., Parry J., Salopek D.S., Stewart J.M., 1994, ApJ, 431, 22}
\smallskip

\quot{Ehlers J., 1993, Gen. Rel. Grav., 25, 1225}
\smallskip

\quot{Ellis G.F.R., 1971, in Sachs R.K., ed., General Relativity and 
Cosmology. Academic Press, New York}
\smallskip

\quot{Ellis G.F.R., 1984, in Bertotti B., de Felice F., \& Pascolini A., eds., 
General Relativity and Gravitation. Reidel, Dordrecht} 
\smallskip

\quot{Ellis G.F.R., 1990, MNRAS, 243, 509} 
\smallskip

\quot{Futamase T., 1988, Phys. Rev. Lett., 61, 2175} 
\smallskip

\quot{Futamase T., 1989, MNRAS, 237, 187} 
\smallskip

\quot{Futamase T., 1991, Prog. Theor. Phys., 86, 389}
\smallskip

\quot{Hawking S.W., Moss I.G., 1982, Phys. Lett., 110B, 35}
\smallskip

\quot{Hui L., Bertschinger E., 1995, preprint astro-ph/9508114} 
\smallskip

\quot{Hwang J.--C., 1994, ApJ, 427, 533}
\smallskip

\quot{Kasai M., 1992, Phys. Rev. Lett., 69, 2330}
\smallskip

\quot{Kasai M., 1993, Phys. Rev., D47, 3214}
\smallskip

\quot{Kasai M., 1995, Phys. Rev., D52, 5605}
\smallskip

\quot{Kellogg O., 1953, Foundations of Potential Theory. Dover, New York}
\smallskip

\quot{Kodama H., Sasaki M., 1984, Prog. Theor. Phys. Suppl., 78, 1}
\smallskip

\quot{Kofman L., Bertschinger E., Gelb M.J., Nusser A., Dekel A., 1994, ApJ, 
420, 44}
\smallskip

\quot{Kofman L., Pogosyan D., 1995, ApJ, 442, 30}
\smallskip

\quot{Icke V., 1973, A\&A, 27, 1}
\smallskip

\quot{Lachi\`eze--Rey M., 1993, ApJ, 408, 403}
\smallskip

\quot{Landau L.D., Lifshitz E.M., 1980, 
The Classical Theory of Fields. Pergamon Press, Oxford} 
\smallskip

\quot{Lesame W. M., Dunsby P.K.S., Ellis G.F.R., 1995, Phys. Rev., D52, 3406} 
\smallskip

\quot{Lesame W.M., Ellis G.F.R., Dunsby P.K.S., 1996, Phys. Rev., D53, 738}
\smallskip

\quot{Lin C.C., Mestel L., Shu F.H., 1965, ApJ, 142, 1431}
\smallskip

\quot{Lynden--Bell D., 1962, Proc. Cambridge Phil. Soc., 58, 709}
\smallskip

\quot{Matarrese S., Pantano O., Saez D., 1993, Phys. Rev., D47, 1311}
\smallskip

\quot{Matarrese S., Pantano O., Saez D., 1994a, Phys. Rev. Lett., 72, 320}
\smallskip

\quot{Matarrese S., Pantano O., Saez D., 1994b, MNRAS, 271, 
513} 
\smallskip

\quot{Moutarde F., Alimi J.-M., Bouchet F.R., Pellat R., Ramani A.,
1991, ApJ, 382, 377}
\smallskip

\quot{Munshi D., Sahni V., Starobinsky A.A., 1994, ApJ, 436, 517}
\smallskip

\quot{Nusser A., Dekel A., 1992, ApJ, 391, 443} 
\smallskip

\quot{Parry J., Salopek D.S., Stewart J.M., 1994, Phys. Rev., D49, 2872}
\smallskip

\quot{Peebles P.J.E., 1980, The Large--Scale Structure of the Universe. 
Princeton University Press, Princeton}
\smallskip

\quot{Peebles P.J.E., 1993, Principles of Physical Cosmology. 
Princeton University Press, Princeton}
\smallskip

\quot{Pyne T., Carroll S.M., 1995, preprint astro-ph/9510041}
\smallskip

\quot{Raychaudhuri A.K., 1957, Z. Astrophys., 43, 161}
\smallskip

\quot{Rees M., Sciama D.W., 1968, Nature, 217, 511}
\smallskip

\quot{Salopek D.S., Stewart J.M., Croudace K.M., 1994, MNRAS, 271, 
1005}
\smallskip

\quot{Schneider P., Ehlers J., Falco E.E., 1992, Gravitational Lenses. 
Springer--Verlag, Berlin}
\smallskip

\quot{Shandarin S.F., Melott A.L., McDavitt K., Pauls J.L., Tinker J., 1995, 
Phys. Rev. Lett., 75, 1}
\smallskip

\quot{Shibata M., Asada H., 1995, Prog. Theor. Phys., 94, 11}
\smallskip

\quot{Szekeres P., 1975, Comm. Math. Phys., 41, 55}
\smallskip

\quot{Thorne K.S., 1995, to appear in Kolb E.W. \& Peccei R., eds., Proc. 
Snowmass 95 Summer Study on Particle and Nuclear Astrophysics and Cosmology. 
World Scientific, Singapore, preprint gr-qc/9506086}
\smallskip

\quot{Tomita K., 1967, Prog. Theor. Phys., 37, 831}
\smallskip

\quot{Tomita K., 1988, Prog. Theor. Phys., 79, 258} 
\smallskip

\quot{Tomita K., 1991, Prog. Theor. Phys., 85, 1041} 
\smallskip

\quot{Weinberg S., 1972, Gravitation and Cosmology. Wiley, New York}
\smallskip

\quot{White S.D.M., Silk J., 1979, ApJ, 231, 1}
\smallskip

\quot{Zel'dovich Ya.B., 1970, A\&A, 5, 84}
\smallskip

\clearpage
\noindent
{\bf Figure 1.} Evolution of the PN tensor modes during the collapse of a 
triaxial homogeneous ellipsoid embedded in an Einstein--de Sitter background;
as well known the generic triaxial case exhibits pancake collapse. 
The {\em top left~} panel shows the behaviour of the physical axes of the 
ellipsoid, $R_A(\tau)=a(\tau) X_A(\tau)$ ($A=1,2,3$), vs. the FRW 
scale--factor $a(\tau)$; the lengths of the three 
axes are initially scaled as $1:1.25:1.5$; the initial density contrast 
is $\delta_0=0.05$; the evolution of the structure constant 
$\alpha_A$ is also shown ({\em top right} panel). 
The {\em bottom left~} panel shows the evolution of the quantities $\mu_A$ 
[defined in Eq.(128)], in terms of which the PN tensor modes $\pi_{AB}$ are 
obtained; 
note that two of the $\mu_A$ tend to diverge near the pancake collapse of the
ellipsoid; as shown in the {\em bottom right} panel, such a divergence is even 
stronger than that of the comoving mass density, $1+\delta(\tau)$.

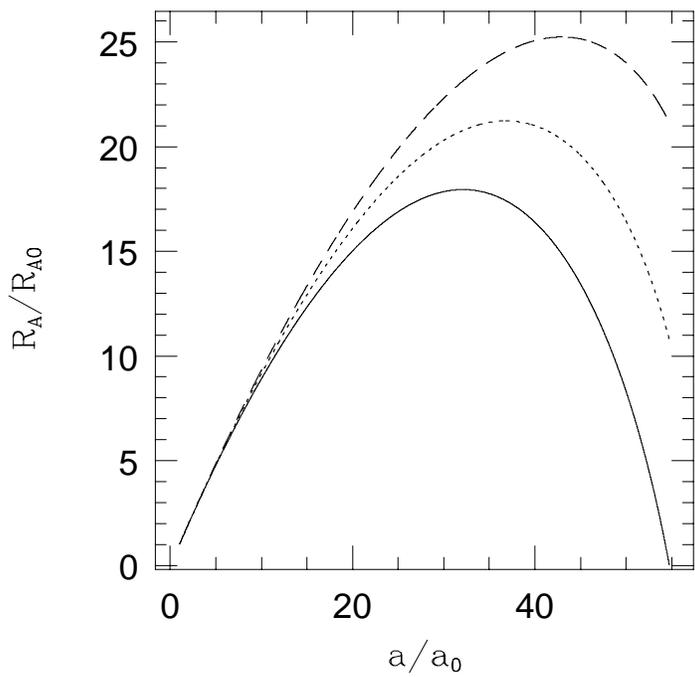
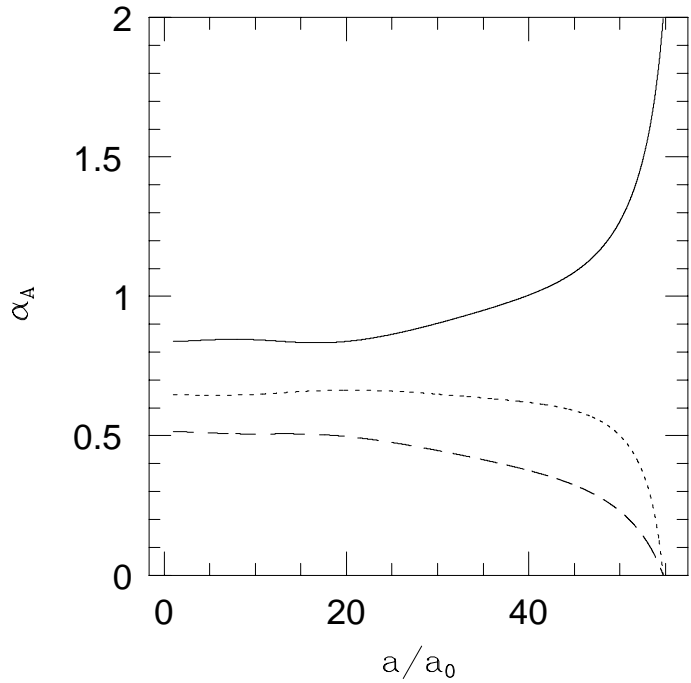
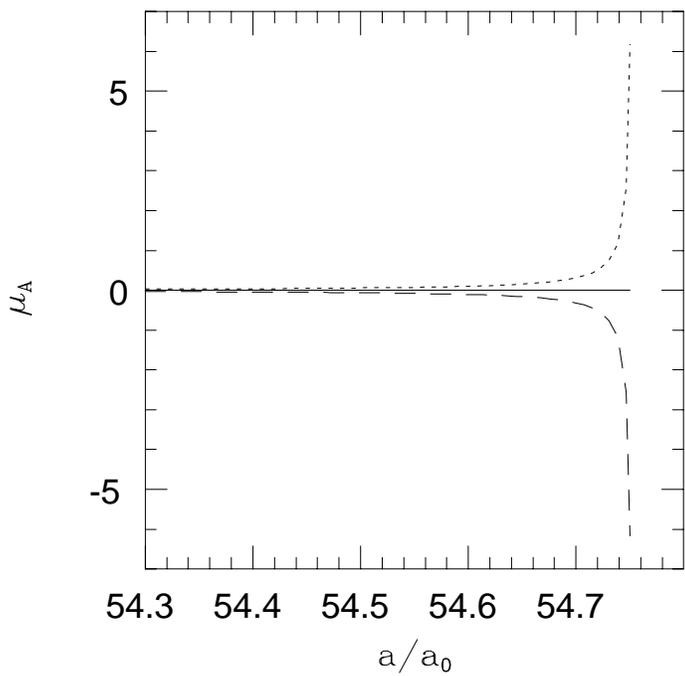
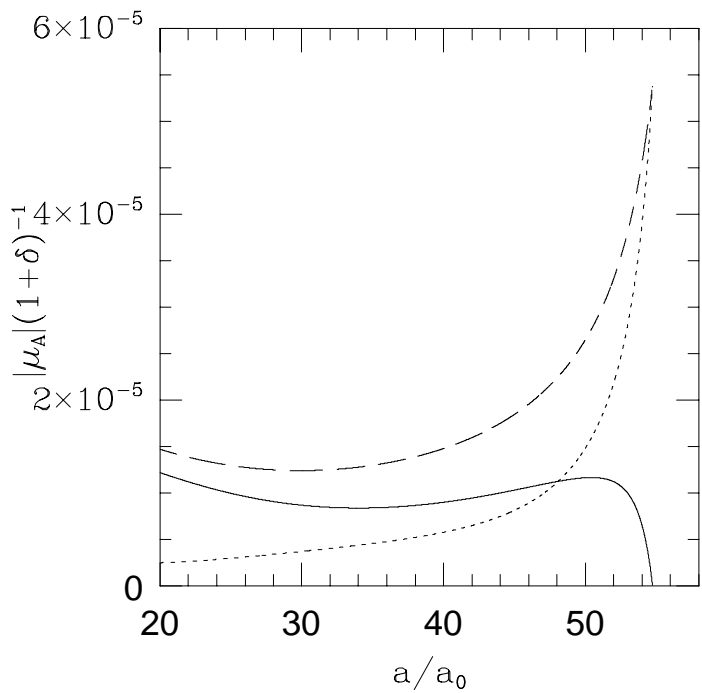
\end{document}